\documentclass[pre,showpacs,twocolumn,preprintnumbers,amsmath,amssymb]{revtex4-1}
\usepackage{epsfig}
\usepackage{graphicx}
\usepackage{textcomp}
\usepackage{epstopdf}
\usepackage{tabularx}
\newcommand{\Op}[1]{{{\mathrm{\hat{#1}}}}}

\begin{document}
\title{Two-strain competition in quasi-neutral stochastic disease dynamics}

\author{Oleg Kogan}
\affiliation{Laboratory of Atomic and Solid State Physics, Cornell University, Ithaca, NY, 14853}
\author{Michael Khasin}
\affiliation{SGT Inc., NASA Ames Research Center, Moffett Field, Mountain View, CA 94035}
\author{Baruch Meerson}
\affiliation{Racah Institute of Physics, Hebrew University of
Jerusalem, Jerusalem 91904, Israel}
\author{David Schneider}
\affiliation{Robert W. Holley Center for Agriculture and Health, Agricultural Research Service, United States Department of Agriculture, and Department of Plant Pathology and Plant-Microbe Biology, Cornell University, Ithaca, NY 14853}
\author{Christopher R. Myers}
\affiliation{Laboratory of Atomic and Solid State Physics, and Institute of Biotechnology, Cornell University, Ithaca, MY 14853}

\begin{abstract}
We develop a new perturbation method for studying quasi-neutral competition in a broad class of stochastic competition models, and apply it to the analysis
of fixation of competing strains in two epidemic models.  The first model is a two-strain generalization of the stochastic Susceptible-Infected-Susceptible (SIS) model.
Here we extend previous results due to Parsons and Quince (2007), Parsons et al (2008) and Lin, Kim and Doering (2012). The second model, a two-strain generalization of the stochastic Susceptible-Infected-Recovered (SIR) model with population turnover, has not been studied previously. In each of the two models, when the basic reproduction numbers of the two strains are identical, a system with an infinite population size approaches a point on the deterministic coexistence line (CL): a straight line of fixed points in the phase space of sub-population sizes.  Shot noise drives one of the strain populations to fixation, and the other to extinction, on a time scale proportional to the total population size.  Our perturbation method explicitly tracks the dynamics of the probability distribution of the sub-populations in the vicinity of the CL.  We argue that, whereas the slow strain has a competitive advantage for mathematically ``typical" initial conditions, it is the fast strain that is more likely to win in the important situation when a few infectives of both strains are introduced into a susceptible population.

\end{abstract}


\maketitle

\section{Introduction}

Competition for resources is a central paradigm in ecology, epidemiology and social sciences \cite{comp1,comp2,comp3}. It is also ubiquitous in physics, for example in the context of mode competition in lasers \cite{lasers}, or the competition for the material among droplets of the minority phase in the process of  Ostwald ripening \cite{Ostwald}.  Competition among different strains of a disease for a pool of susceptible individuals (or for resources of a single individual infected with multiple strains) arises naturally in outbreaks of infectious diseases. This is because diseases commonly occur in multiple strains that result from mutations \cite{Keeling}.  Citing a recent review \cite{Balmer}, ``multiple-strain infections have been shown unambiguously in 51 human pathogens (and 21 non-human ones) and are likely to arise in most pathogen species.... Competition and mutualism between
strains change pathogen and disease dynamics and promote pathogen evolution." It is, therefore, important to understand how different strains compete among themselves. One way of achieving this goal is to use mathematical models
of spread of infectious diseases in populations.

Two basic models of this type, and their extensions, have been especially popular: the Susceptible-Infected-Susceptible (SIS) and the Susceptible-Infected-Recovered (SIR) model \cite{Keeling,Brauer,SISandSI}.  In the SIS model an individual can be in either of the two states, and can transit from the susceptible to infected state upon a contact with another infected, or recover and become susceptible again.  In the SIR model with population turnover, an infected individual can be removed (leave, recover with immunity or die), while the susceptibles are removed and renewed.  The basic reproduction number $\mathcal{R}_0$, which is the average number of new infectives produced by an infected host in a fully susceptible population, is given in both of these models by the rate of infectivity divided by the rate of recovery.
These two simple models have been extended in different directions \cite{Keeling,Brauer,SISandSI}. One direction is to incorporate multiple strains of disease \cite{Parsons1,Karrer,Grassberger}.  This paper continues this line of research by studying stochastic competition between two strains in the context of the SIS model and the SIR model with
population turnover. We focus on the important special case when both strains have identical
values of $\mathcal{R}_0$.  In this case the rate of infectivity of the first strain is a fraction of the rate of infectivity of the second strain, and the rate of recovery of the first strain is the same fraction of the rate of recovery of the second strain.  Thus, one strain may be labeled as the ``faster'', and the other as the ``slower''.  This situation with different rates but identical $\mathcal{R}_0$ has been termed ``quasi-neutral'', since the strains are only neutral in
the deterministic limit \cite{Parsons1, Parsons2}.  We refer the reader to Ref. \cite{Parsons1} for evolutionary arguments for importance of the quasi-neutral case. In the quasi-neutral case small fluctuations due to the finite size of the population (which we call the shot noise) have a much stronger impact on the long-time behavior of the disease, than in the cases with different $\mathcal{R}_0$.  In the deterministic limit
each of the two models exhibits a coexistence line (CL), which is a line of fixed points.  The shot noise qualitatively changes the dynamics - the system effectively performs random walk along the CL, resulting ultimately in the extinction of one of the strains, usually referred to as fixation of the other strain.  The mean time for fixation scales with the characteristic
population size $N$ \cite{Parsons2}.

The previous works on quasi-neutral competition \cite{Parsons1,Parsons2,Doering} dealt with two-population models systems. Here we develop a new perturbation method that can be applied to multi-population models.  The method is based on time scale separation.

We use this perturbation method to study fixation in two-strain extensions of the SIS and SIR models: the  $\text{SI}_{\text{1}}\text{I}_{\text{2}}\text{S}$ model and the $\text{SI}_{\text{1}}\text{I}_{\text{2}}\text{R}$ model with population turnover. A model, mathematically identical to the  $\text{SI}_{\text{1}}\text{I}_{\text{2}}\text{S}$ model, was studied earlier,  using a different method, in the context of population genetics
\cite{Parsons1, Parsons2}, see also Ref. \cite{Doering}. As the total population size in the $\text{SI}_{\text{1}}\text{I}_{\text{2}}\text{S}$ model is fixed, this model is two-dimensional and simpler for analysis. For pedagogical reasons, we will introduce our perturbation method for the $\text{SI}_{\text{1}}\text{I}_{\text{2}}\text{S}$ model. 
As in Ref. \cite{Doering}, we derive an effectively one-dimensional description of the dynamics of the probability distribution along the CL. Apart from that, we provide an explicit description of the dynamics of the  probability distribution of the sub-populations in the vicinity of the CL. Our method is not limited to two-dimensional models, as we demonstrate for the intrinsically three-dimensional stochastic $\text{SI}_{\text{1}}\text{I}_{\text{2}}\text{R}$ model with population turnover. Here we reduce the three-dimensional model to effectively one-dimensional and determine  analytically the fixation probability and the mean time to fixation.

We also show, for both models, that the competitive advantage of strains depends in a somewhat peculiar way on the initial conditions.  For a uniform distribution of initial conditions the slow strain is more likely to reach fixation, as observed earlier in models with a fixed total population size \cite{Parsons1, Parsons2, Doering}.
The fast strain, however, is more likely to win the competition in the important situation when only a few infectives of both strains are introduced into a susceptible population.

In Section \ref{sec:Model} we introduce the   $\text{SI}_{\text{1}}\text{I}_{\text{2}}\text{S}$ model and the $\text{SI}_{\text{1}}\text{I}_{\text{2}}\text{R}$ models with population turnover and discuss the nature of their deterministic solutions focusing on the quasi-neutral case. Our perturbation method is presented in Section \ref{method} that starts with a qualitative discussion of how the shot noise leads to extinction of one strain and fixation of the other. It also discusses the time scale separation that is crucial to the perturbation method. The derivation itself is presented
in Section \ref{sec:2d_calculation} for the simpler $\text{SI}_{\text{1}}\text{I}_{\text{2}}\text{S}$ model.  We then apply the method, in Section \ref{sec:3d_calculation}, to the more involved $\text{SI}_{\text{1}}\text{I}_{\text{2}}\text{R}$ model with population turnover.  We compute the fixation probabilities and the mean fixation times for both models in Section \ref{sec:1d_solution}. Section \ref{sec:CompAdv} analyzes the competitive advantage of the strains for different initial conditions.  The results are summarized and discussed in Section \ref{sec:Discussion}.

\section{Models}
\label{sec:Model}
\subsection{$\text{SI}_{\text{1}}\text{I}_{\text{2}}\text{S}$ model and its deterministic limit}
Consider two infectious strains competing for the same susceptible population in the framework of the SIS model \cite{Keeling,Brauer}.  This model includes the following processes, see Table \ref{table1}. A susceptible individual $S$ can become $I_1$, that is infected with strain 1, upon contact with another $I_1$.  The rate
of this process is $(\beta_1/N) SI_1$, where $N\gg 1$ is a fixed total population size.  An $I_1$ can recover with rate $\kappa_1 I_1$ and become susceptible again.  The same two processes occur for strain 2, except that the rates are now $(\beta_2/N) S I_2$ and $\kappa_2I_2$, respectively.

\begingroup
\squeezetable
\begin{table}[ht]
\begin{ruledtabular}
\begin{tabular}{|c|c|c|}
 Event & Type of transition &  Rate\\
  \hline
  Infection with strain 1& $S\to S-1, \, I_1\to I_1+1$ &  $(\beta_1/N) SI_1$\\
  Infection with strain 2& $S\to S-1, \, I_2\to I_2+1$ &  $(\beta_2/N) SI_2$\\
  Recovery of $I_1$ & $I_1\to I_1-1, \, S\to S+1$ & $\kappa_1 I_1$ \\
  Recovery of $I_2$& $I_2\to I_2-1, \, S\to S+1$ & $\kappa_2 I_2$ \\
\end{tabular}
\end{ruledtabular}
\caption{Transition rates for the stochastic $S I_1I_2S$ model}\label{table1}
\end{table}
\endgroup

Let us start with the deterministic limit of this model. Introducing the basic reproduction numbers $\mathcal{R}_1 =\beta_1/\kappa_1$ and $\mathcal{R}_2 = \beta_2/\kappa_2$ for $I_1$ and $I_2$ respectively,  we can write the deterministic equations:
\begin{eqnarray}
\label{eq:MF-original}
\dot{I}_1 &=& \kappa_1\left(\frac{\mathcal{R}_1}{N}  S - 1\right)I_1, \nonumber\\
\dot{I}_2 &=& \kappa_2\left(\frac{\mathcal{R}_2}{N}  S - 1\right)I_2, \nonumber\\
\dot{S}  &=& -\kappa_1\left(\frac{\mathcal{R}_1}{N}  S - 1\right)I_1 - \kappa_2\left(\frac{\mathcal{R}_2}{N}  S - 1\right)I_2,
\end{eqnarray}
where the dots stand for the time derivatives.  Since the total population size $I_1+I_2+S=N=\text{const}$, we can eliminate $S$ and obtain
\begin{eqnarray}
\label{eq:MF-originala}
\dot{I}_1 &=& \kappa_1\left[\mathcal{R}_1-1-\frac{\mathcal{R}_1}{N}  (I_1+I_2)\right] I_1, \nonumber\\
\dot{I}_2 &=& \kappa_2\left[\mathcal{R}_2-1-\frac{\mathcal{R}_2}{N}  (I_1+I_2)\right] I_2,
\end{eqnarray}
so the system is two-dimensional. We assume that $\mathcal{R}_1>1$ and $\mathcal{R}_2>1$. If, in addition, $\mathcal{R}_1 \neq \mathcal{R}_2$, the dynamical system (\ref{eq:MF-originala}) has three fixed points (FPs) with non-negative population sizes: FP1 at $\left[I_1 = 0, I_2 = N\left(1-1/\mathcal{R}_2\right)\right]$, FP2 at $\left[I_1 = N\left(1-1/\mathcal{R}_1\right), I_2 = 0\right]$, and FP3 at $\left(I_1 = 0, I_2 = 0\right)$.   FP3 is a repeller; of the other two fixed points one is a saddle, and the other is an attractor.  Both $I_1=0$ and $I_2 = 0$ lines are absorbing: the  system can not escape from them.  If $\mathcal{R}_1 > \mathcal{R}_2$, FP1 is a saddle point: the attracting eigenvector lies along the $I_2$ axis, and the repelling eigenvector has a non-zero $I_1$ component,
while FP2 has both eigenvectors attracting.  If $\mathcal{R}_1 < \mathcal{R}_2$, the character of the fixed points FP1 and FP2 is interchanged. Therefore, the state with the larger $\mathcal{R}_i$ (a one-strain state) is the only attracting state:  no coexistence of the two strains is possible. For any initial condition with nonzero $I_1$ and $I_2$, the system approaches the attracting fixed point. The characteristic relaxation time scale -- the time scale for reaching the vicinity of the globally attracting fixed point -- is determined by the eigenvalues of the attracting state and is independent of the total population size $N$.

A different picture emerges in the particular case of our interest here: $\mathcal{R}_1 = \mathcal{R}_2 \equiv \mathcal{R}$.  Here the non-trivial fixed points of the dynamical system obey the relation
\begin{equation}
I_1 + I_2= N\left(1-\frac{1}{\mathcal{R}}\right),
\end{equation}
and form a straight line -- the deterministic \emph{coexistence line} (CL) -- on the $I_1I_2$ plane.
This CL describes a continuum family of endemic states with an arbitrary proportion of strains 1 and 2,
whereas the edge points $N(1-1/\mathcal{R},0)$ and $N(0,1-1/\mathcal{R})$ of the CL describe one-strain endemic states.

Introducing the rescaled population sizes $x=I_1/N$ and $y=I_2/N$, denoting $a = \kappa_2/\kappa_1$ and rescaling time by $1/\kappa_1$, we can rewrite the deterministic equations as
\begin{eqnarray}
\label{eq:mf2d}
\dot{x} &=&\mathcal{R}x(1-x-y) - x, \nonumber \\
\dot{y} &=&a\left[\mathcal{R}y(1-x-y)-y\right],
\end{eqnarray}
whereas the CL is described by the equation
\begin{equation}
x+y=
1-1/\mathcal{R} \equiv r.
\end{equation}
Equations~(\ref{eq:mf2d}) have two rescaled parameters:  the basic reproduction number  $\mathcal{R} >1 $, or $0 < r < 1$, and the parameter $a>0$. Without losing generality we can assume $a\leq 1$.
When $\mathcal{R}>1$, the system approaches the CL on a fast, $N$-independent relaxation time scale  along one of the lines
\begin{equation}
\label{phaseplane1}
y =  M x^a\,,
\end{equation}
\begin{figure}[ht]
\includegraphics[width=3in]{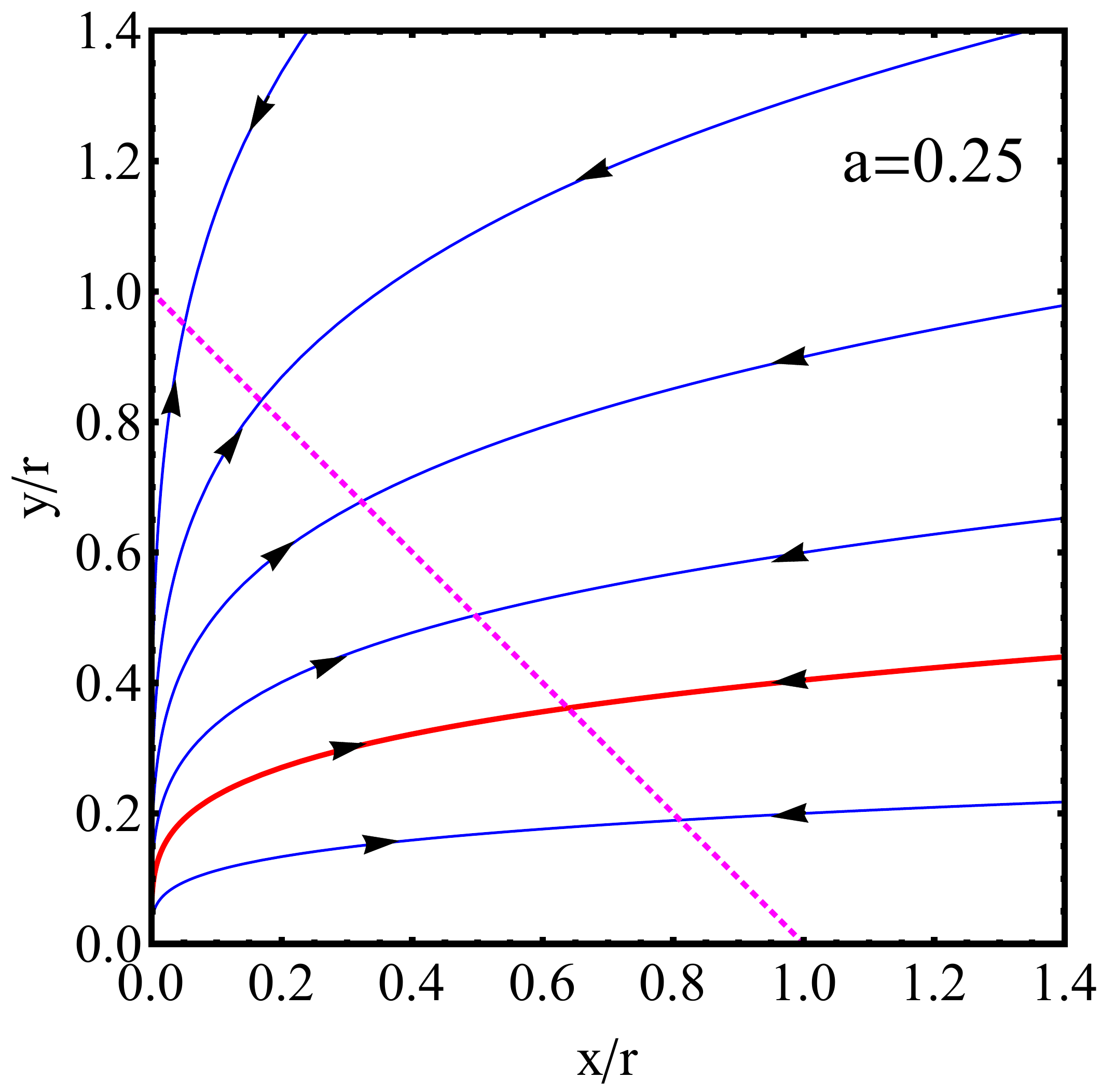}
\caption{(Color online) The deterministic phase plane of the quasi-neutral $\text{SI}_{\text{1}}\text{I}_{\text{2}}\text{S}$ model in the rescaled variables $x/r=I_1/(r N)$ and $y/r=I_2/(r N)$, where $r=1-1/\mathcal{R}$. The coexistence line (CL) is indicated by the dotted line. The thick curve
is the borderline of the phase diagram of the stochastic quasi-neutral system: The
fixation probability of each of the strains is equal to $1/2$ on this curve. For the points of the CL below (above) the borderline the fixation of strain 1 is more (less) likely than that of strain 2. The parameter $a=0.25$.}
\label{fig:0}
\end{figure}
\\
parameterized by $0\le M < \infty$. Equation~(\ref{phaseplane1}) can be obtained by dividing the second of Eq.~(\ref{eq:mf2d}) by the first one and integrating the resulting differential equation for $y=y(x)$.
The arbitrary constant $M$ is set by the initial conditions $x(t=0)$ and $y(t=0)$.
Figure \ref{fig:0} shows the phase plane of the system.

\subsection{$\text{SI}_{\text{1}}\text{I}_{\text{2}}\text{R}$ model with population turnover and its deterministic limit}

Our second model deals with two infectious strains competing for the same susceptible population in the framework of the SIR model (where R stands for Recovered or Removed) with population turnover \cite{Keeling,Brauer,SISandSI}. The SIR model includes the following processes, see Table \ref{table2}.  The susceptibles $S$ are removed (leave or die) with rate $\mu^{\prime} S$ and renewed with constant rate $\mu^{\prime} N$, where the large parameter $N\gg 1$ sets the scale of population size. A susceptible individual $S$ becomes $I_1$, that is infected with strain 1 with rate $(\beta_1/N) SI_1$.  An $I_1$ is removed - leaves, recovers with immunity or dies -- with rate  $\kappa_1 I_1$.  The same two processes occur for strain 2, except that the rate constants are now $(\beta_2/N) S I_2$ and $\kappa_2 I_2$, respectively.

\begingroup
\squeezetable
\begin{table}[ht]
\begin{ruledtabular}
\begin{tabular}{|c|c|c|}
 Event & Type of transition &  Rate\\
  \hline
  Removal of susceptibles& $S\to S-1$ &  $ \mu^{\prime} S$\\
  Renewal of susceptibles& $S\to S+1$ &  $ \mu^{\prime} N $\\
  Infection with strain 1& $S\to S-1, \, I_1\to I_1+1$ &  $(\beta_1/N) SI_1$\\
  Infection with strain 2& $S\to S-1, \, I_2\to I_2+1$ &  $(\beta_2/N) SI_2$\\
  Removal of $I_1$& $I_1\to I_1-1$ & $\kappa_1 I_1$ \\
  Removal of $I_2$& $I_2\to I_2-1$
  & $\kappa_2 I_2$ \\
\end{tabular}
\end{ruledtabular}
\caption{Transition rates for the stochastic $\text{SI}_{\text{1}}\text{I}_{\text{2}}\text{R}$ model with population turnover}\label{table2}
\end{table}
\endgroup

Introducing the basic reproduction numbers $\mathcal{R}_1 \equiv\beta_1/\kappa_1$ and $\mathcal{R}_2 \equiv \beta_2/\kappa_2$ for $I_1$ and $I_2$ respectively, we can write the deterministic equations for $I_1$, $I_2$ and $S$:
\begin{eqnarray}
\label{eq:MF-original1}
\dot{I_1} &=& \kappa_1\left(\frac{\mathcal{R}_1}{N}  S - 1\right)I_1, \nonumber\\
\dot{I_2} &=& \kappa_2\left(\frac{\mathcal{R}_2}{N}  S - 1\right)I_2, \nonumber\\
\dot{S}  &=& \mu^{\prime} (N-S) - \frac{\mathcal{R}_1 \kappa_1}{N} I_1 S - \frac{\mathcal{R}_2 \kappa_2}{N} I_2 S,
\end{eqnarray}
This model is intrinsically three-dimensional \cite{removed}.
For ${\mathcal R}_1>1$, ${\mathcal R}_2>1$ and ${\mathcal R}_1\neq {\mathcal R}_2$, the dynamical system (\ref{eq:MF-original1}) has three fixed points with non-negative populations: FP1 at $\left[I_1 = 0, I_2 = \frac{\mu N}{\kappa_2}\left(1 - \frac{1}{\mathcal{R}_1}\right), S=\frac{N}{\mathcal{R}_2}\right]$, FP2 at $\left[I_1 = \frac{\mu N}{\kappa_1} \left(1-\frac{1}{\mathcal{R}_1}\right), I_2 = 0, S=\frac{N}{\mathcal{R}_1}\right]$ and FP3 at $\left(I_1 = 0, I_2 = 0, S = N\right)$. FP3 is a saddle with only one attracting direction: the one along the $S$ axis.
Of the other two fixed points, one is a saddle, the other is an attractor.  Both $I_1=0$ and $I_2 = 0$ planes are absorbing.  If $\mathcal{R}_1 > \mathcal{R}_2$, FP1 is a saddle, with two attracting directions in the $I_2S$ plane, and the third direction coming out of this plane is repelling.
FP2 has all three directions that are attracting.  If $\mathcal{R}_1 < \mathcal{R}_2$ the character of the fixed points FP1 and FP2 is interchanged.  Thus, as in the $\text{SI}_{\text{1}}\text{I}_{\text{2}}\text{S}$ model with ${\mathcal R}_1\neq {\mathcal R}_2$, there is always only one globally attracting state.  The time scale for reaching the vicinity of this globally-attracting state, the relaxation time scale, is determined by the eigenvalues of the attracting state and is independent of $N$.

When $\mathcal{R}_1 = \mathcal{R}_2 \equiv \mathcal{R}$, the non-trivial fixed points of this dynamical system obey the relations
\begin{equation}
\kappa_1 I_1 + \kappa_2 I_2=\mu^{\prime} N\left(1-\frac{1}{\mathcal{R}}\right)\,,\;\;\;S=\frac{N}{\mathcal{R}}\,
\end{equation}
and form a straight line: the deterministic coexistence line (CL) in the three-dimensional phase space $I_1I_2S$.

\begin{figure}[ht]
\includegraphics[width=2.5in]{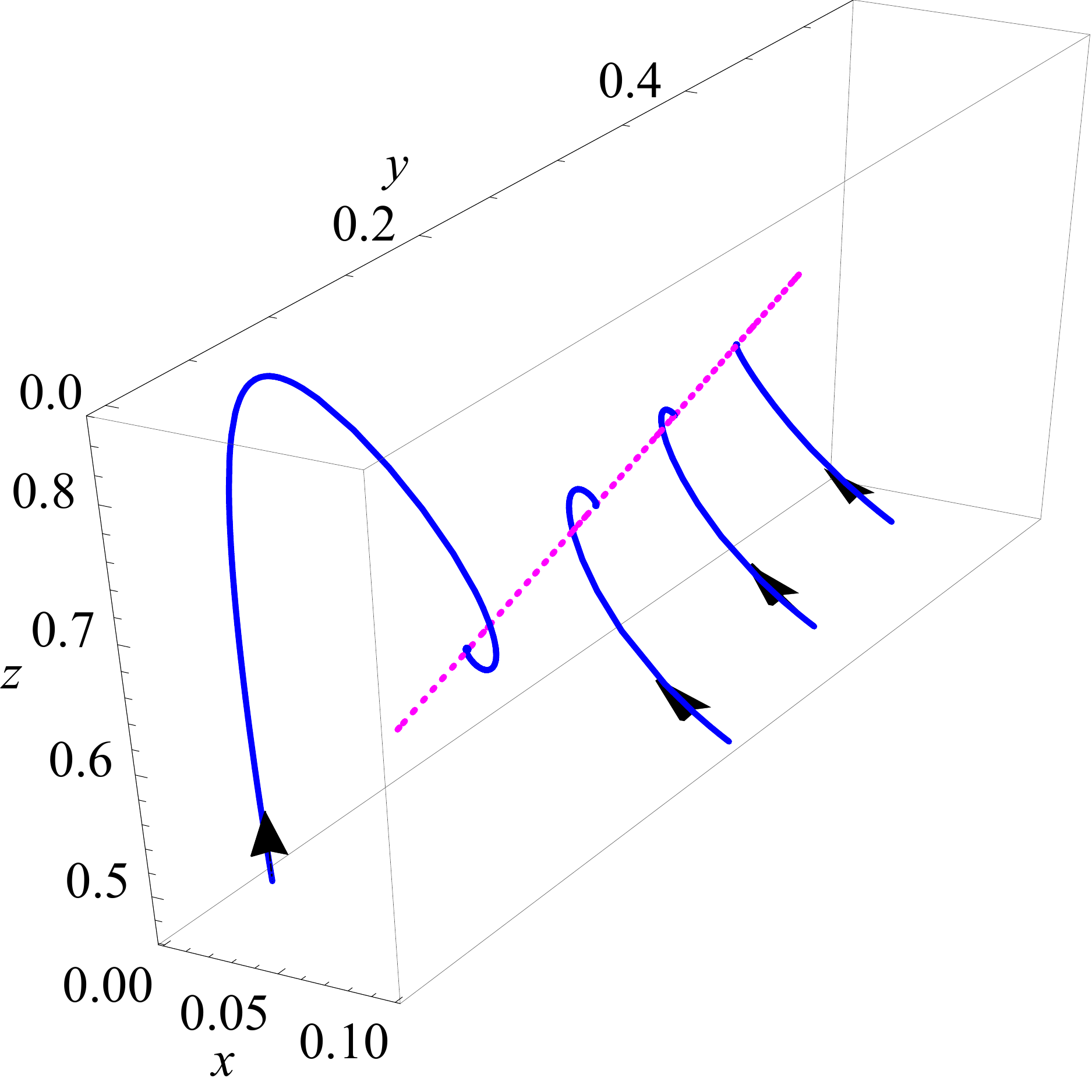}
\caption{(Color online) A sketch of the deterministic phase space of the $\text{SI}_{\text{1}}\text{I}_{\text{2}}\text{R}$ model in the rescaled variables $x=I_1/N$, $y=I_2/N$ and $z=S/N$. The coexistence line (CL) is shown by the dotted line. The rescaled parameters are ${\mathcal R}=1.5$, $\mu=0.3$ and $a=0.2$.}
\label{fig:100}
\end{figure}

As in the $\text{SI}_{\text{1}}\text{I}_{\text{2}}\text{S}$ model, when $\mathcal{R}>1$, the deterministic trajectories approach the CL on the fast relaxation time scale, independent of $N\gg 1$.  In contrast to the $\text{SI}_{\text{1}}\text{I}_{\text{2}}\text{S}$ model, here the character of fixed points making the CL can change depending on the parameters ${\mathcal R}$, $\mu$ and $a$ and, in general, on the coordinate along the CL. A point of the CL can be either a stable node, or a stable spiral in the direction transverse to the CL, see Appendix \ref{sec:MFAppendix}. Introducing the rescaled population sizes $x=I_1/N$, $y=I_2/N$ and $z=S/N$, denoting $a = \kappa_2/\kappa_1\leq 1$ and $\mu = \mu^{\prime}/\kappa_1$, and rescaling time by $1/\kappa_1$ we can rewrite the deterministic equations (\ref{eq:MF-original1}) for the $\text{SI}_{\text{1}}\text{I}_{\text{2}}\text{R}$ model with population turnover as
\begin{eqnarray}
\label{eq:mf}
\dot x&=& x(\mathcal{R} z-1)\,, \nonumber \\
\dot y&=&a y (\mathcal{R} z-1)\,,\nonumber \\
\dot z&=&\mu(1- z) - \mathcal{R} z(x+a y)\,,
\end{eqnarray}
whereas the rescaled CL is given by
\begin{equation}
\label{rescaledCL}
x+a y=r\equiv \mu \left(1-1/\mathcal{R}\right)\,,\;\;\;z=1/\mathcal{R}\,.
\end{equation}
Unlike in the case of the $\text{SI}_{\text{1}}\text{I}_{\text{2}}\text{S}$ model, here the length of the CL increases as $a$ becomes smaller.
As one can see, the deterministic theory of the $\text{SI}_{\text{1}}\text{I}_{\text{2}}\text{R}$ model is characterized by three rescaled parameters:  the basic reproduction number $1 \leq \mathcal{R} \leq \infty$, or $0 \leq r \leq 1$, the rescaled rate constant $\mu$ and the parameter $a$. Figure \ref{fig:100} shows a sketch of the deterministic phase space of the quasi-neutral $\text{SI}_{\text{1}}\text{I}_{\text{2}}\text{R}$ model in the rescaled variables $x$, $y$ and $z$.

Dividing the second of Eqs.~(\ref{eq:mf}) by the first one and integrating, we obtain the equation
\begin{equation}
\label{phaseplane}
y =  M x^a\,,\;\;\;0\le M < \infty,
\end{equation}
which coincides with Eq.~(\ref{phaseplane1}). That is, the \emph{projections} of the phase space trajectories of the $\text{SI}_{\text{1}}\text{I}_{\text{2}}\text{R}$ model onto the $I_1I_2$ plane lie on curves that coincide with the phase trajectories of the $\text{SI}_{\text{1}}\text{I}_{\text{2}}\text{S}$ model, see Fig.~\ref{fig:0}. This property  holds for a whole family of quasi-neutral competition models described by the rescaled equations of the type
\begin{eqnarray*}
  \dot{x} &=& x \,{\mathcal K}(x,y,z), \nonumber \\
  \dot{y} &=& a y \, {\mathcal K}(x,y,z), 
\end{eqnarray*}
and an equation for $\dot{z}$, leading to Eq.~(\ref{phaseplane}).

\section{Perturbation method and effective one-dimensional Fokker-Planck equation}
\label{method}

\subsection{Quasi-neutral stochastic dynamics: a qualitative picture and time scale separation}
\label{sec:approach_discussion}

\begin{figure}[ht]
\includegraphics[width=0.45\textwidth,clip=]{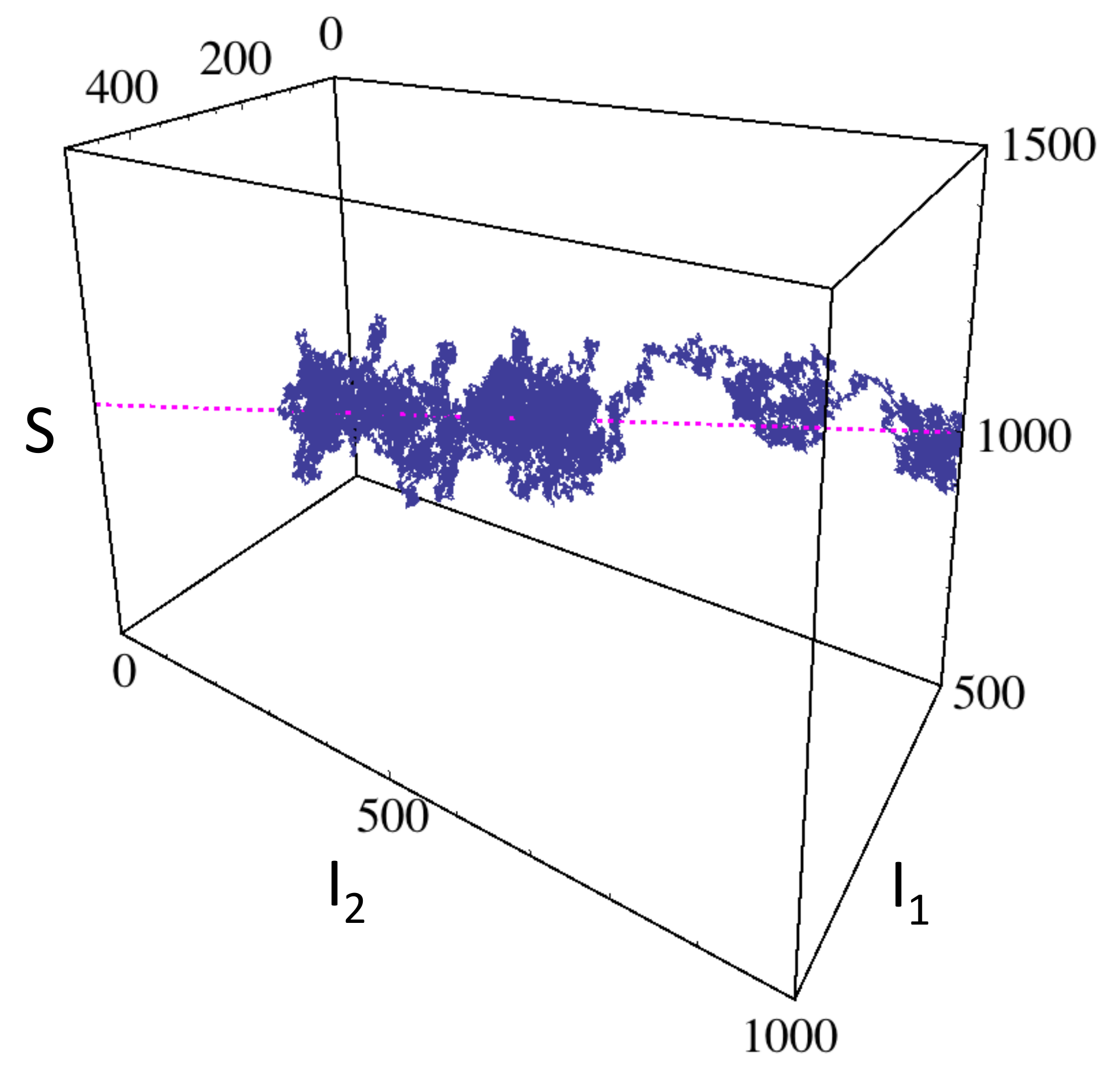}
\includegraphics[width=0.25\textwidth,clip=]{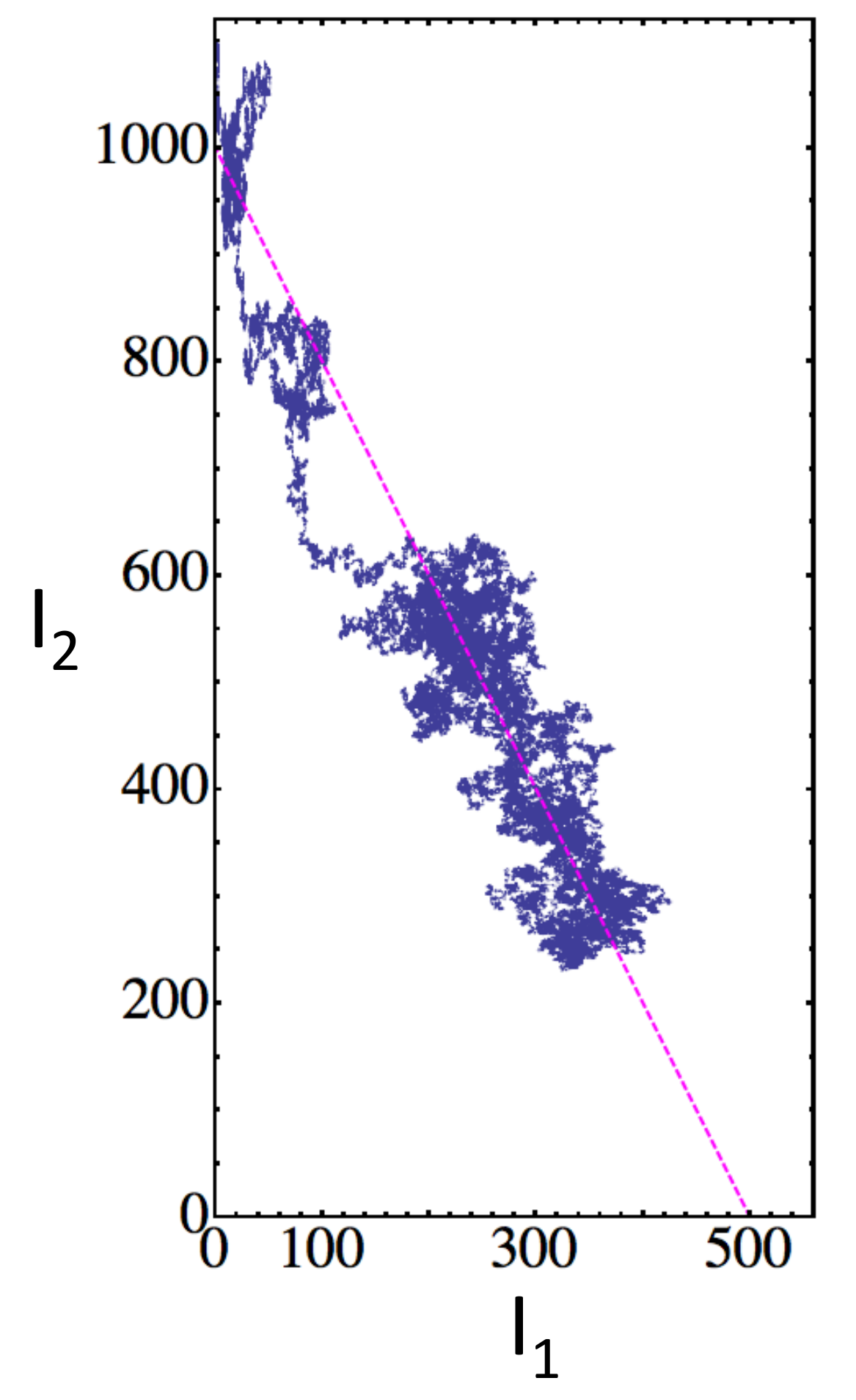}
\caption{(Color online) A stochastic realization of the $\text{SI}_{\text{1}}\text{I}_{\text{2}}\text{R}$ model with population turnover
in the phase space of $I_1, I_2$ and $S$ (the upper panel), and its projection onto the $I_1, I_2$ plane (the lower panel). Extinction of $I_1$ and fixation of $I_2$ can be seen. The deterministic CL and its projection onto the $I_1, I_2$ plane are shown as dashed lines. The parameters are $R = 2$, $a=0.5$, $\mu=0.5$ and $N=2000$.}
\label{fig:2}
\end{figure}

Before embarking on the derivation of the perturbation method, we give a physical picture and a road map we follow in the remainder of the paper.
The random character of elementary processes of infection, recovery, etc. introduces shot noise into the system.
In the quasi-neutral case the shot noise  changes qualitatively the nature of the dynamics compared to predictions from the deterministic theory. This is because the noise makes the system wander randomly (mostly) along the CL, eventually reaching extinction of one strain and fixation of the other \cite{Parsons1,Parsons2,Doering}.  This effect is illustrated by a sample stochastic trajectory of the $\text{SI}_{\text{1}}\text{I}_{\text{2}}\text{R}$ model, generated using Gillespie algorithm \cite{Gillespie} and shown in Fig.~\ref{fig:2}.

At the level of probabilistic description, we characterize the system by the probability distribution to have, at time $t$,
certain population sizes of each relevant sub-population. For example, for the $\text{SI}_{\text{1}}\text{I}_{\text{2}}\text{R}$ model, which we will use for explanations in this subsection, this probability distribution is $P_{m,n}(t)$, where $m\geq 0$ and $n\geq 0$ denote the population sizes of strains 1 and 2, respectively. The time evolution of the probability distribution is described by the master equations presented below: for the $\text{SI}_{\text{1}}\text{I}_{\text{2}}\text{S}$ model [Eq.~(\ref{eq:master_master2d})] and for the $\text{SI}_{\text{1}}\text{I}_{\text{2}}\text{R}$ model [Eq.~(\ref{eq:MasterEquation})]. Employing the van Kampen system size expansion, based on the small parameter $1/N\ll 1$, we will first approximate the master equation by a Fokker-Planck equation of corresponding dimension \cite{Gardiner}. Then we will employ time-scale separation, intrinsic to
the quasi-neutral stochastic dynamics, and derive an effective \emph{one-dimensional} Fokker-Planck equation for the slowly-evolving probability distribution of the system along the CL.  This one-dimensional Fokker-Planck equation then can be analyzed in a standard way \cite{Gardiner} to
determine the fixation probabilities and the mean time to fixation of each of the two strains.

Throughout this work we assume $N \gg 1$. We also assume a ``macroscopic'' initial condition $P(m,n,t=0)=\delta_{m,m_0} \,\delta_{n,n_0}$ that involves fixed (and sufficiently large) numbers of infected with the two strains: $m_0, n_0 \gg 1$.  In this case the evolution of the probability distribution $P(m,n,t)$  has three distinct stages. During the first stage,  $P_{m,n}(t)$ develops a sharp peak at the CL around the stable fixed point that is determined by  $x_0 = m_0/N$ and $y_0=n_0/N$: the (rescaled) initial numbers of infected with strains 1 and 2.  The characteristic formation time  of this peaked distribution is independent of $N$ and therefore short.

During the much slower second stage (which duration turns out to be $\sim N$), this sharp peak evolves into a sharp ridge, as the probability distribution spreads along the CL.  It is the probability distribution spread along the CL that ultimately causes the extinction of one strain and fixation of the other upon reaching the end of the CL at $m = 0$ or $n = 0$.   Throughout this process, large fluctuations away from the CL are suppressed by the deterministic drift toward the CL.
The with of the sharp ridge around the CL, where noise and the deterministic flow are comparable, is $\sim1/\sqrt{N}$.

The still much longer third stage involves an \emph{exponentially} slow leakage of the single-strain probability
distribution to the infection-free state, leading to a complete extinction of the disease from the populations. The
extinction of a single-strain endemic disease has been extensively studied for the SIS model with and without population turnover \cite{SISextinction}, and for the SIR model with population turnover \cite{KM,SIextinction}. The mean time of the disease extinction here is exponentially large in $N$.
In this work we are interested in the intermediate second stage that determines which of the two strains has a competitive advantage to become established, for a very long time, in the susceptible population.

\subsection{$\text{SI}_{\text{1}}\text{I}_{\text{2}}\text{S}$ model: a case of two dimensions}
\label{sec:2d_calculation}
The Markov stochastic dynamics in the discrete state space of the sub-population sizes is described by the master equation for the probability $P_{m,n}(t)$ to observe $m$ individuals infected with strain $1$ and $n$ individuals infected with strain $2$.  With time rescaled by $1/\mu$, as in the deterministic equations~(\ref{eq:mf2d}),  this master equation is
\small
\begin{multline}
\label{eq:master_master2d}
\dot P_{m,n}(t) = \\
\shoveleft{\frac{\mathcal{R}}{N}(m-1)(N-m+1-n)P_{m-1,n}(t) -\frac{\mathcal{R}}{N}m(N-m-n)P_{m,n}(t)} \\
\shoveleft{+(m+1)P_{m+1,n}(t)-mP_{m,n}(t)} \\
\shoveleft{+a\frac{\mathcal{R}}{N}(n-1)(N-m+1-n)P_{m,n-1}(t)-a\frac{\mathcal{R}}{N}n(N-m-n)P_{m,n}(t)} \\
+a(n+1)P_{m,n+1}(t)-a n P_{m,n}(t),
\end{multline}
\normalsize
Using the large parameter $N\gg 1$, we can perform the van Kampen system size expansion \cite{Gardiner} and approximate the exact master equation (\ref{eq:master_master2d}) by the Fokker-Planck equation for the quasi-continuous probability density $\rho(x,y,t)$:
\begin{eqnarray}
\label{eq:fp2d}
\partial_t \rho(x,y,t)&=& -\frac{\partial}{\partial x}\left\{\left[R x (1-x-y)-x\right]\rho \right\}\nonumber \\
&-&a \frac{\partial}{\partial y}\left\{\left[R y (1-x-y)-y\right]\rho \right\} \nonumber \\
&+&\frac{1}{2N}\frac{\partial^2}{\partial x^2}\left\{\left[R x (1-x-y)+x\right]\rho \right\}\nonumber \\ &+&\frac{a}{2N}\frac{\partial^2}{\partial y^2}\left\{\left[R y (1-x-y)+y\right]\rho \right\}.
\end{eqnarray}
The small noise enters the equation via the diffusion terms that scale as $1/N\ll 1$. We anticipate that the noise rapidly establishes a sharp distribution across the CL and then slowly spreads this distribution along the CL.  Let us introduce the new variables
\begin{eqnarray}
X &=& x-y \nonumber \\
Y^{\prime} &=& x+y-r,
\end{eqnarray}
where $X$ is the slow variable that measures the distance along the CL, and $Y^{\prime} $ is the fast variable that measures the distance away from the CL.  The CL is given by $Y^{\prime}=0$, so $x = (r+X)/2$ and $y = (r-X)/2$ on the CL.  In the new variables the Fokker-Planck equation is
\small
\begin{multline}
\label{eq:fp2d_2}
\partial_t \rho(X,Y^{\prime},t)= \\
\shoveleft{-\left(\frac{\partial}{\partial Y^{\prime}}+\frac{\partial}{\partial X}\right)\left\{\frac{Y^{\prime}+X +r}{2}\left[\mathcal{R} (1-Y^{\prime}-r)-1\right]\rho \right\}} \\
\shoveleft{-a\left(\frac{\partial}{\partial Y^{\prime}}-\frac{\partial}{\partial X}\right)\left\{\frac{Y^{\prime}-X +r}{2}\left[\mathcal{R} (1-Y^{\prime}-r)-1\right]\rho \right\}} \\
\shoveleft{+\frac{1}{2N}\left(\frac{\partial}{\partial Y^{\prime}}+\frac{\partial}{\partial X}\right)^2 \left\{\frac{Y^{\prime}+X + r}{2}\left[\mathcal{R} (1-Y^{\prime}-r)+1\right]\rho \right\}} \\
\shoveleft{+\frac{a}{2N}\left(\frac{\partial}{\partial Y^{\prime}}-\frac{\partial}{\partial X}\right)^2 \left\{\frac{Y^{\prime}-X + r}{2}\left[\mathcal{R} (1-Y^{\prime}-r)+1\right]\rho \right\}.}
\end{multline}
\normalsize
Since we expect the distribution of the fast variable $Y^{\prime}$ to be sharply peaked about $Y^{\prime}=0$, with a characteristic width $\sim 1/\sqrt{N}$, we introduce the new variable $Y = \sqrt{N} Y^{\prime}$.  Now $Y \sim 1$ in the region of the ridge where the probability density $\rho(X,Y,t)$ is substantial.  Expanding the right hand side of Eq.~(\ref{eq:fp2d_2}) in powers of the small parameter $\varepsilon = 1/\sqrt{N}$ up to the second order, we obtain
\begin{equation}
\label{eq:exp2d}
\partial_t \rho(X,Y,t) = \left(\Op L^{(0)}+\varepsilon \Op L^{(1)}+ \varepsilon ^2 \Op L^{(2)}\right)\rho (X,Y,t),
\end{equation}
where the linear differential operators $\Op L^{(0)}$, $\Op L^{(1)}$ and $\Op L^{(2)}$  are presented in Appendix \ref{sec:Operators_appendix_2d}. Importantly, the operator  $\Op L^{(0)}$ involves differentiation only with respect to the fast variable $Y$.  Being interested in the solution of Eq.~(\ref{eq:exp2d}) that develops on a slow time scale of ${\mathcal O}(\varepsilon^{-2})= \mathcal O(N)$, we make the ansatz
\begin{eqnarray}
\label{ansatz}
\rho(X,Y,t) &=& \rho^{(0)}(X, Y, \varepsilon^2 t)+\varepsilon \rho^{(1)}(X,Y,\varepsilon^2 t) \nonumber \\
&+& \varepsilon^2 \rho^{(2)}(X,Y,\varepsilon^2 t)+ \ldots .
\end{eqnarray}
Plugging it into Eq.~(\ref{eq:exp2d}) we obtain
\begin{equation}
\label{eq:zero}
\Op L^{(0)} \rho^{(0)}=0
\end{equation}
in the zeroth order of $\varepsilon$,
\begin{equation}
\label{eq:first}
\Op L^{(0)}\rho^{(1)}=-\Op L^{(1)}\rho^{(0)}
\end{equation}
in the first order of $\varepsilon$, and
\begin{equation}
\label{eq:second}
\Op L^{(0)}\rho^{(2)}=\partial_{\tau} \rho^{(0)} - \Op L^{(1)}\rho^{(1)}-\Op L^{(2)}\rho^{(0)}
\end{equation}
in the second order of $\varepsilon$.  Here $\tau=\varepsilon^2 t=t/N$ is the slow time.
The solution to Eq.~(\ref{eq:zero})
can be written as
\begin{equation}
\label{eq:zerosol}
\rho^{(0)}(X,Y, \tau)=\sqrt{\frac{\mathcal{R}}{2\pi}} \, f(X,\tau) \,e^{-\frac{\mathcal{R}}{2}Y^2},
\end{equation}
where $f(X,\tau)$ is an arbitrary function. The function $\rho^{(0)}(X,Y,\tau)$, with yet unknown $f(X,\tau)$, is
the ``ridge distribution" announced above. It is a Gaussian with respect to $Y$, that is a sharp Gaussian of width $\sim N^{-1/2}$ with respect to $Y^{\prime}$.

The slow temporal evolution of $\rho^{(0)}$, i.e.~of the $Y$-independent function $f(X,\tau)$ is described by Eq.~(\ref{eq:second}).  To obtain an evolution equation for  $f(X,\tau)$,
we can integrate Eq.~(\ref{eq:second}) with respect to $Y$ from $-\infty$ to $\infty$. Since the left hand side of Eq.~(\ref{eq:second}) is a full derivative with respect to $Y$ [see Eqs.~(\ref{zero}) and~(\ref{geneq:adjoint_structure2})], it vanishes upon the integration, and we obtain
\begin{equation}
\label{eq:secondb}
 \int_{-\infty}^{\infty} \partial_{\tau}\rho^{(0)} \, dY=   \partial_{\tau}f = \int_{-\infty}^{\infty} \left(\Op L^{(1)}\rho^{(1)}+\Op L^{(2)}\rho^{(0)}\right)\,dY.
\end{equation}
The  integration of the second term on the right hand side of Eq.~(\ref{eq:secondb}) reduces to the computation of the zero and second moments of the Gaussian distribution (\ref{eq:zerosol}) and can be performed right away (see Appendix \ref{sec:Operators_appendix_2d} for the explicit forms of the operators):
\begin{eqnarray}
&&\int_{-\infty}^{\infty}\Op L^{(2)}\rho^{(0)} d Y = \frac{1-a}{2}\partial_X f+\frac{1}{2}\partial^2_{X}\left[h(X)f\right], \nonumber \\
&&h(X)=(1+a)r+(1-a)X\,. \label{eq:nextorder10}
\end{eqnarray}
It remains to compute the integral of the first term on the r.h.s. of Eq.~(\ref{eq:secondb}). A straightforward way to proceed would be to first find $\rho^{(1)}$ from Eq.~(\ref{eq:first}) that arises in the first order in $\varepsilon$. Although this is not
hard to do in two-dimensional models like the $\text{SI}_{\text{1}}\text{I}_{\text{2}}\text{S}$, the solution becomes difficult, if at all feasible, in higher dimensions. Fortunately, a bypass is possible for a whole class of quasi-neutral competition models. The key idea is to avoid solving for $\rho^{(1)}$, by exchanging it for $\rho^{(0)}$.  As we explain in Appendix \ref{bypass}, it can be done with the help of a function $F(X,Y)$ such that
\begin{equation}
\label{eq:key1}
\int \Op L^{(1)} \rho^{(1)} \,dY \equiv \partial_X\int F(X,Y) \Op L^{(0)} \rho^{(1)} \,dY .
\end{equation}
Then, using Eq.~(\ref{eq:first}), we obtain
\begin{equation}
\label{eq:key2}
\int \Op L^{(1)} \rho^{(1)} \,dY =- \partial_X\int F(X,Y) \Op L^{(1)} \rho^{(0)} \,dY.
\end{equation}
We show in Appendix \ref{sec:application} how to calculate the function $F(X,Y)$. The result is:
\begin{eqnarray}
&&F(X,Y)=-\frac{g(X)}{h(X)}Y, \nonumber \\
&&g(X)=(1-a)r+(1+a)X\,, \label{eq:Fsol}
\end{eqnarray}
and $h(X)$ was defined in Eq.~(\ref{eq:nextorder10}). Now Eq.~(\ref{eq:key2}) becomes
\begin{multline}
\label{term2a}
\int_{-\infty}^{\infty}\Op L^{(1)}\rho^{(1)}(X,Y) dY =\\  -\partial_X\left[\frac{(1+a)g(X)}{h(X)}f(X) + \frac{g^2(X)}{2h(X)} \partial_X f \right].
\end{multline}
The details of computing the integral on the r.h.s. of Eq.~(\ref{eq:key2}) are given in Eqs.~(\ref{eq:IBP})-(\ref{eq:2dL1result}).

Adding up the two terms, Eqs.~(\ref{eq:nextorder10}) and (\ref{term2a}), in the r.h.s. of Eq.~(\ref{eq:secondb}), we finally arrive at an effective one-dimensional Fokker-Planck equation
\begin{eqnarray}
\label{eq:nextorder11}
  &&\partial_t f(X,t)\nonumber \\
   &&=   \frac{2a(1-a)}{N}\,\frac{\partial}{\partial X} \left\{\frac{1-(X/r)^2}{\left[1+a+(X/r) (1-a)\right]^2}f\right\} \nonumber\\
  && +\frac{2ar}{N}\,\frac{\partial^2}{\partial X^2}\left[\frac{1-(X/r)^2}{1+a+(X/r) (1-a)}f\right].
\end{eqnarray}
This equation describes an effective Markov process along the CL \cite{Doering}.  It involves slow drift and diffusion, both $X$-dependent.  Noticeable is the same scaling behavior $\sim 1/N$ of the drift and diffusion coefficients.  Not only the diffusion, but the drift as well is induced by the shot noise. The drift introduces a bias in favor of the slow strain, for which $a<1$. For $a=1$, when the two strains are identical, the drift term vanishes, and one is left with $X$-dependent diffusion
coefficient that is symmetric with respect to $X$. In this particular case Eq.~(\ref{eq:nextorder11}) coincides with (the diffusion approximation of) the Moran model, a minimalist model of random genetic variations in a haploid population, see e.g. Ref. \cite{BlytheMcKane}.
As expected, the drift and diffusion coefficients both vanish at the absorbing boundaries $X=\pm r$ signaling extinction of one strain and fixation of the other.

The physical mechanism of the noise-induced drift becomes clear if we consider
the following schematic model. Let at some instant of time the system is at a point $x_0, y_0$ on the CL. Then the system
is ``kicked", in a small time interval $dt$, by the shot noise to a new point $x_1=x_0+\rho_0 \cos \phi$, $y_1=y_0+\rho_0 \sin \phi $, where $\rho_0>0$
and  $\phi$ are independent random variables. Crucially, we assume the kicks to be isotropic: $\phi$ is uniformly distributed on the interval $0\leq\phi<2\pi$. The distribution of $\rho$ is less important, it should have  a proper variance $\sim 1/N$. Following the kick, the system returns to the CL along the deterministic trajectory passing through $x_1, y_1$.  Lin et al \cite{Doering}
\begin{figure}[ht]
\includegraphics[width=2.1in]{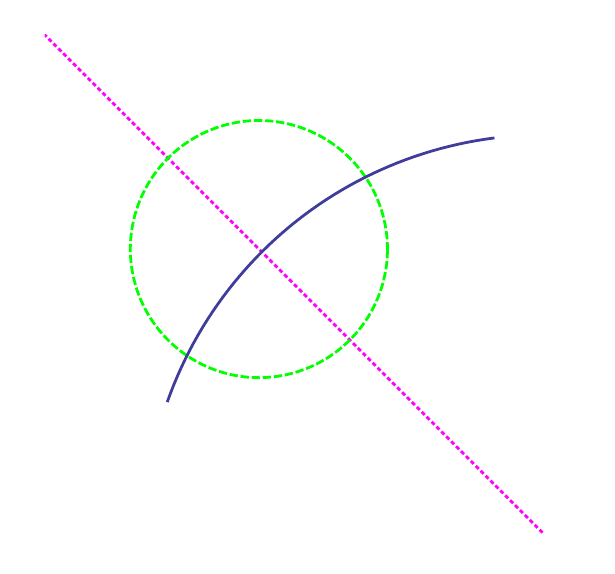}
\caption{(Color online) The noise-induced drift (the ratchet effect), favoring the slow strain, is caused by an interplay of the shot noise and the curvature of the deterministic phase trajectories, see the text.  Shown are segments of the CL (the dotted line) and of the deterministic phase trajectory (\ref{phaseplane1}) at $a<1$ (the solid line), passing through the point $x_0, y_0$ on the CL.}
\label{fig:curvature}
\end{figure}
employed a qualitatively similar schematic model for calculating the effective drift and diffusion coefficients in a quasi-neutral competition model
mathematically identical to the $\text{SI}_{\text{1}}\text{I}_{\text{2}}\text{S}$ model, and their results are identical to those following from our Eqs.~(\ref{eq:nextorder11}). Here we only use this model to elucidate the origin of the drift and its sign, see Fig.~\ref{fig:curvature}.  Because of the curvature of the deterministic phase trajectories
(\ref{phaseplane1}), it is more likely that the system will return to a point on the CL that is to the left of the original point (that is closer to the fixation point of the slow strain) than to the right, see Fig.~\ref{fig:curvature}. It is this ``geometric ratchet" that, when combined with the isotropic shot noise, introduces a systematic drift favoring the slower competitor.  The drift coefficient is proportional in this picture to the effective variance of the shot noise, that is to $1/N$. Importantly, the geometric ratchet mechanism is independent of the specific competition model. When $a=1$, the phase trajectories (\ref{phaseplane1}) becomes straight lines. In this case the ``geometric ratchet" effect is absent and drift coefficient vanishes.

Now we return to Eq.~(\ref{eq:nextorder11}). It is convenient to rescale the coordinate along the CL by introducing $\xi=X/r$, such that $\xi \in [-1,1]$.  When $\xi = -1$, $I_1 = 0$, and when $\xi = 1$, $I_2 = 0$.  After an additional rescaling of time, $\tau= t/(N r)$, we can rewrite Eq.~(\ref{eq:nextorder11}) in a universal form, independent of $r$:
\begin{equation}
\label{eq:FP-final}
\frac{\partial f(\xi,\tau)}{\partial \tau}=-\frac{\partial}{\partial \xi}\left[v(\xi) f(\xi,\tau)\right]+\frac{1}{2}\frac{\partial^2}{\partial \xi^2}\left[D(\xi) f(\xi,\tau)\right]\,,
\end{equation}
where the effective (and rescaled) drift and diffusion coefficients are
\begin{equation}\label{AB1}
v(\xi) =-\frac{2a(1-a) (1-\xi^2)}{[1+a+(1-a)\xi]^2}
\end{equation}
and
\begin{equation}\label{AB2}
D(\xi) = \frac{4a(1-\xi^2)}{1+a+(1-a)\xi}\,,
\end{equation}
respectively.  The effective initial condition for Eq.~(\ref{eq:FP-final}) is
\begin{equation}
\label{in}
f(\xi,0)=\delta(\xi-\xi_0)\,,
\end{equation}
where $\xi_0$ is determined by $x_0=m_0/N$ and $y_0=n_0/N$: the (rescaled) initial  numbers of infected with strains 1 and 2.  Namely, to determine $\xi_0$ we should find
the intersection point of the phase trajectory [see Eq.~(\ref{phaseplane})]
\begin{equation}\label{x0y0}
\frac{y}{y_0}=\left(\frac{x}{x_0}\right)^a
\end{equation}
and the CL $x+y=r$, and transform to the coordinate $\xi$. We shall analyze the effective one-dimensional
problem in the next Section. Here its worth emphasizing that, once the effective problem is solved and $f(X,t)$ found, our method gives an explicit description of the dynamics of the probability distribution of the sub-populations in the vicinity of the CL. In the leading order in $1/\sqrt{N}$, this description is given by Eq.~(\ref{eq:zerosol}).

\subsection{$\text{SI}_{\text{1}}\text{I}_{\text{2}}\text{R}$ model with population turnover: a case of three dimensions}
\label{sec:3d_calculation}
The $\text{SI}_{\text{1}}\text{I}_{\text{2}}\text{R}$ model with population turnover is more involved, as it is
fully three-dimensional. An application of our perturbation method here leads again to an effective one-dimensional Fokker-Planck equation along the CL. We start with the master equation:
\begin{eqnarray}
\label{eq:MasterEquation}
 && \dot{P}_{m,n,s}(t) \nonumber \\
  && =\frac{\mathcal{R}}{N}(m-1)(s+1)P_{m-1, n, s+1}(t) - \frac{\mathcal{R}}{N}msP_{m,n,s}(t) \nonumber \\
  && + a\frac{\mathcal{R}}{N}(n-1)(s+1)P_{m,n-1,s+1}(t) - a\frac{\mathcal{R}}{N}nsP_{m,n,s}(t) \nonumber\\
  && +(m+1)P_{m+1,n,s}(t) - m P_{m,n,s}(t) \nonumber\\
  && +a (n+1)P_{m,n+1,s}(t) - a n P_{m,n,s}(t) \nonumber\\
  && +\mu N P_{m,n,s-1}(t) - \mu N P_{m,n,s}(t) \nonumber \\
  && +\mu (s+1) P_{m, n,s+1}(t) - \mu s P_{m,n,s}(t)
\end{eqnarray}
that describes the time evolution of probability $P_{m,n,s}(t)$ to observe $m$ individuals infected with strain $1$, $n$ individuals infected with strain $2$, and $s$ susceptible individuals.  Time $t$ has been rescaled by $1/\kappa_1$.  We perform the van Kampen system size expansion and switch to the new coordinates: $X$ along the CL, and $Y$ and $Z$  perpendicular to the CL. As in the $\text{SI}_{\text{1}}\text{I}_{\text{2}}\text{S}$ model, we rescale the perpendicular coordinates by $\sqrt{N}$:
\begin{eqnarray}
\label{eq:newvarnoN}
X &=& -ax+y , \nonumber \\
Y &=& \sqrt{N}\left(x+ay-r\right), \nonumber\\
Z &=& \sqrt{N}\left(z-\frac{1}{\mathcal{R}}\right).
\end{eqnarray}
The CL is determined by Eqs.~(\ref{rescaledCL});  as one can check, $x = (r-aX)/(1+a^2)$ and $y = (X+ar)/(1+a^2)$ on the CL.  After some algebra, the resulting three-dimensional Fokker-Planck equation can again be presented in the form of Eq.~(\ref{eq:exp2d}), with operators $\Op L^{(n)}$, $n=0,1,2$, presented in Appendix \ref{sec:Operators_appendix_3d}.  Now we make the perturbation ansatz~(\ref{ansatz1}) and define the slow time $\tau = t/N$.  Putting all this together and collecting orders of $\varepsilon$, we arrive at the same three operator equations~(\ref{eq:zero})-(\ref{eq:second}) as before.

The solution of Eq.~(\ref{eq:zero}) is a bivariate Gaussian distribution:
\small
\begin{equation}
\label{eq:rho^0}
\rho^{(0)}(X,Y,Z) = \mathcal{N} f(X,\tau)\, e^{A(X)Y^2 +  B(X) YZ + C(X)Z^2},
\end{equation}
\normalsize
where $\mathcal{N}$ is the normalization factor with respect to $Y$ and $Z$ variables.
In contrast to the two-dimensional case, the coefficients $A$, $B$ and $C$ are generally $X$-dependent.
Using the ansatz~(\ref{eq:rho^0}), we find
\begin{eqnarray}
\label{eq:Asoln}
A(X) &=& \Re(X)B(X), \\
\label{eq:Bsoln}
B(X) &=& -\left[4\Re^2(X) c_0(X) + 2\Re(X)d_0(X) + \mu\right]^{-1}, \\
\label{eq:Csoln}
C(X) &=& \!-\frac{c_0(X) B(X)}{2\mu^2 \mathcal{R}} - \frac{\mathcal{R}}{2},\\
\Re(X) &=& \frac{1}{2}\left[\frac{1}{\mu \mathcal{R}}-\frac{d_0(X)}{c_0(X)}\right],
\end{eqnarray}
and the expressions for $c_0(X)$ and $d_0(X)$ are given in Appendix \ref{sec:Operators_appendix_3d}.
Diagonalizing the quadratic form in $Y$ and $Z$ in the exponential in Eq.~(\ref{eq:rho^0}), we obtain the following expression in terms of the principal coordinates $\chi$ and $\zeta$:
\begin{equation}
\label{eq:diagonalized_Gaussian}
\rho^{(0)}(X,Y,Z) = \mathcal{N} f(X,\tau)\,e^{\Lambda_+(X) \chi^2 + \Lambda_-(X) \zeta^2},
\end{equation}
where
\begin{equation}
\label{eq:Lambdapm}
\Lambda_{\pm} (X)= \frac{1}{2}\left[(A+C) \pm \sqrt{(A-C)^2 + B^2}\right]
\end{equation}
are negative real numbers.
In terms of $\Lambda_{\pm}$, the normalization factor of the bi-Gaussian distribution is
\begin{equation}
\mathcal{N} = \pi^{-1} \sqrt{\Lambda_+(X)\Lambda_-(X)}.
\end{equation}
The principal directions are given by
\begin{equation}
\label{eq:Vpm}
\vec{V}_{\pm} =  \frac{1}{n_{\pm}} \left(\begin{array}{c} 1 \\ \\ \frac{2(\Lambda_{\pm}-A)}{B} \end{array}\right)
\end{equation}
where $n_{\pm}$ is a normalization factor.
It is worth mentioning that the principal directions of the bi-Gaussian distribution do not coincide with the attracting eignevectors of the CL in the deterministic theory,   see Appendix \ref{sec:MFAppendix}.

In a full analogy with Eq.~(\ref{eq:secondb}) for the $\text{SI}_{\text{1}}\text{I}_{\text{2}}\text{S}$ model, the evolution of the slow variable distribution is now given by
\small
\begin{equation}
\label{eq:solvability_condition}
\partial_{\tau} f(X,\tau)  = \int_{-\infty}^{\infty} \hat{L}^{(1)}\rho^{(1)} dYdZ+ \int_{-\infty}^{\infty} \hat{L}^{(2)}\rho^{(0)}dYdZ.
\end{equation}
\normalsize
Here and in the following
$$
\int_{-\infty}^{\infty} \ldots dY dZ
$$
denotes integration over both $Y$ and $Z$ from $-\infty$ to $\infty$. Calculation of the second term on the right hand side of Eq.~(\ref{eq:solvability_condition}) reduces to the calculation of the zeroth and second moments of the bi-Gaussian distribution (\ref{eq:diagonalized_Gaussian}):
\begin{eqnarray}
\label{eq:second_term}
&&\int_{-\infty}^{\infty} \hat{L}^{(2)}\rho^{(0)}dYdZ \nonumber \\
&&= \partial_X\left\{\frac{a\mathcal{R}(1-a)}{1+a^2}\langle YZ \rangle f  +  a\partial_X\left[\frac{2ar + (1-a^2)X}{1+a^2}f \right]\right\} \nonumber \\
\end{eqnarray}
where
\begin{eqnarray}
\langle YZ \rangle &=& \frac{q}{2(1+q^2)}\left(\frac{1}{|\Lambda_+|} - \frac{1}{|\Lambda_-|}\right), \;\;\;
\mbox{and} \nonumber\\
q(X) &=& \frac{2[\Lambda_+(X) - A(X)]}{B(X)}. \label{eq:second_moment}
\end{eqnarray}

Calculation of the first term in Eq.~(\ref{eq:solvability_condition}) boils down to finding the function $F(X,Y,Z)$, such that
\begin{equation}
\int \Op L^{(1)} \rho^{(1)} \,dY \ dZ=- \partial_{X}\int F(X,Y,Z) \Op L^{(1)} \rho^{(0)} \,dY \ dZ. \label{eq:key2b}
\end{equation}
similarly to what was done for the $\text{SI}_{\text{1}}\text{I}_{\text{2}}\text{S}$ model and following the general procedure outlined in Appendix \ref{bypass}.
The solution (see Appendix \ref{sec:application}) is:
\begin{eqnarray}
&&F (X,Y,Z)= -\frac{\Xi(X)}{\mathcal{R} d_0(X)}Y, \nonumber\\
&&\Xi(X)=\frac{\mathcal{R}a \left[(1-a)r - (1+a)X\right]}{1+a^2}, \label{eq:F3d}
\end{eqnarray}
and $d_0(X)$ is given in Eq.~(\ref{eq:c-d-coeffs}).  In its turn, Eq.~(\ref{eq:key2b}) leads to
\begin{eqnarray}
\label{eq:adjoint_structureb2}
&&\int_{-\infty}^{\infty}\Op L^{(1)}\rho^{(1)} dYdZ \nonumber \\
&&=-\partial_X\left\{\left[\frac{2aG(X)}{1+a^2}\partial_X f  - \frac{\mathcal{R}ap(X)}{1+a^2}\partial_X \left(\langle YZ \rangle f\right) \right. \right.\nonumber \\
&&+ \left. \left. \frac{4a^2 - \mathcal{R}(1-a-a^2+a^3)\langle YZ \rangle}{1+a^2}f
 \right]\frac{ap(X)}{(1+a^2)d_0(X)} \right\} \nonumber \\
\end{eqnarray}
where
\begin{eqnarray}
&&G(X)=2aX - (1-a^2)r, \\
&&p(X)=(1-a)r - (1+a)X.
\end{eqnarray}
and $\langle YZ \rangle $ is given by Eq.~(\ref{eq:second_moment}) above.

Finally, we rescale the coordinate along the CL
\begin{equation}
\xi = 1 - \frac{2a}{r} \frac{X+ar}{1+a^2} = -\frac{G(X)}{(1+a^2)r},
\end{equation}
such that $\xi \in [-1,1]$.  When $\xi = -1$, we have $I_1 = 0$, and when $\xi = 1$, we have $I_2 = 0$, as in the $\text{SI}_{\text{1}}\text{I}_{\text{2}}\text{S}$ model. Inserting Eqs.~(\ref{eq:second_term}) and (\ref{eq:adjoint_structureb2}) into Eq.~(\ref{eq:solvability_condition}), making simplifications and expressing the resulting equation in the variable $\xi$ and a newly rescaled time $\tau = t/(Nr)$ in the standard Fokker-Planck form, we again arrive at Eq.~(\ref{eq:FP-final}) with the effective (and rescaled) drift and diffusion coefficients
\begin{equation}
\label{eq:v3d}
v(\xi) = -\frac{2a(1-a)(1 -\xi^2)  \left[1+(1-a^2)\xi+ a(a+4)\right]}{[1+a + (1-a)\xi]^3}
\end{equation}
and
\begin{equation}
\label{eq:D3d}
D(\xi) = \frac{8a^2(1-\xi^2)}{[1+a + (1-a)\xi]^2},
\end{equation}
respectively.  The effective initial condition for Eq.~(\ref{eq:FP-final}) is again Eq.~(\ref{in}), where $\xi_0$ is fully determined by $x_0=m/N$ and $y_0=n/N$: the rescaled initial numbers of infected with strains 1 and 2. To determine $\xi_0$ we should find $x$ and $y$ from  Eq.~(\ref{x0y0}) and the CL equation $x+ay=r$, and then transform to the coordinate $\xi$. Because of the degeneracy, intrinsic to the quasi-neutral competition, $\xi_0$ is independent of the initial number of susceptibles $z_0$.

We shall analyze the effective one-dimensional problem in the next Section. As in the $\text{SI}_{\text{1}}\text{I}_{\text{2}}\text{S}$, once the one-dimensional problem is solved, the method gives
an explicit description of the dynamics of the probability distribution of the sub-populations in the vicinity of the CL.  In the leading order in $1/\sqrt{N}$ this description is provided by Eq.~(\ref{eq:rho^0}).

\section{Effective dynamics along the CL}
\label{sec:1d_solution}
\begin{figure}[ht]
\includegraphics[width=3.5in,clip=]{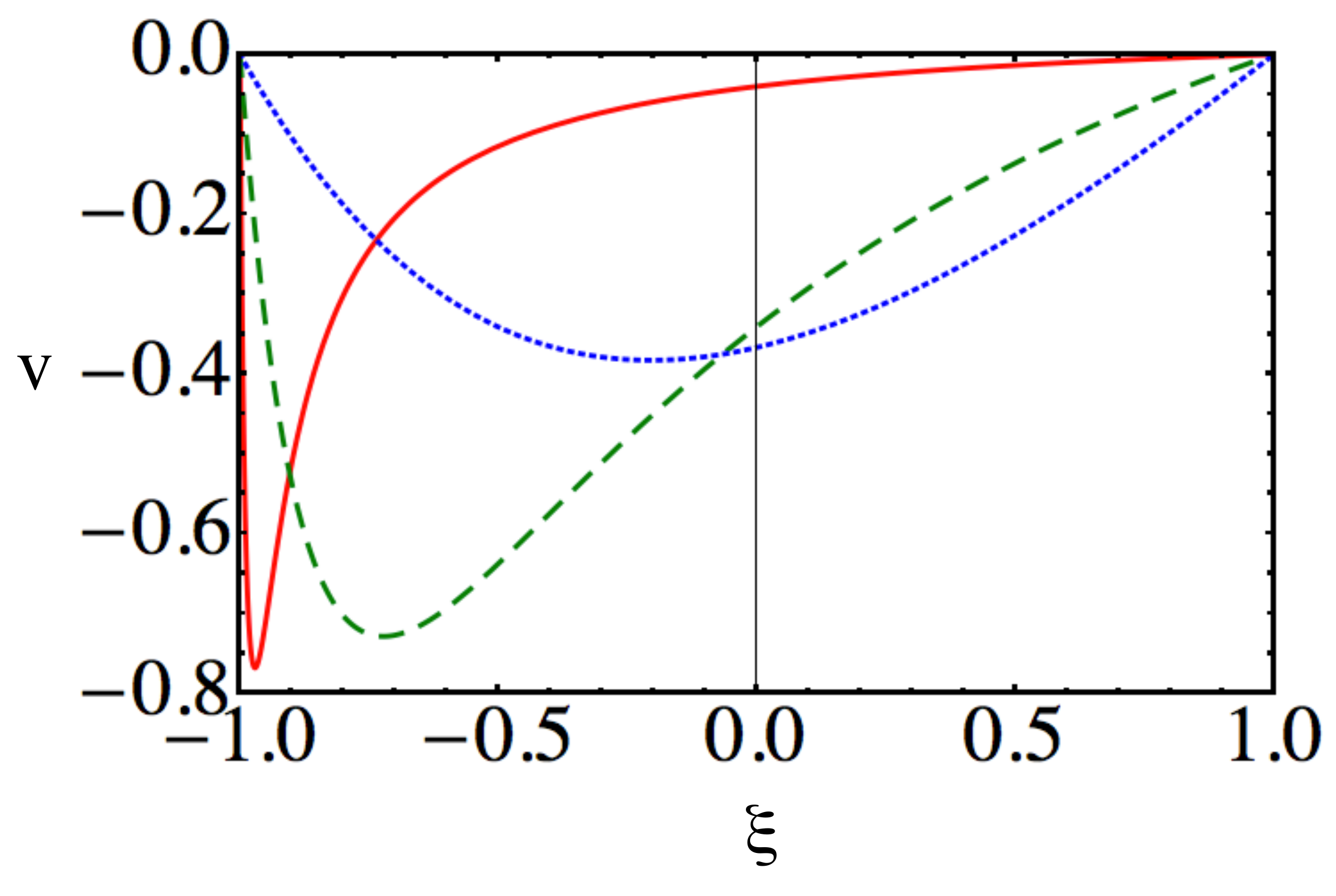}
\caption{(Color online) Effective drift coefficient along the coexistence line, $v(\xi)$   from Eq.~(\ref{eq:v3d}), for the  $\text{SI}_{\text{1}}\text{I}_{\text{2}}\text{R}$  model with $a=0.02$ (solid line), $a=0.2$ (dashed line), and $a=0.7$ (dotted line).   As $a \rightarrow 0$, $v(\xi)$ becomes strongly localized at $\xi=-1$, whereas the minimum value $v_{\text{min}}(\xi)$ approaches $v_{min} = -4/(3\sqrt{3}) = - 0.7698\ldots$. The drift vanishes at $a=1$.}
\label{fig:vplot}
\end{figure}

We now employ the effective one-dimensional evolution equations that we derived to study the quasi-neutral competition. Several examples of the $\xi$-dependence of the drift coefficient $v(\xi)$ for the  $\text{SI}_{\text{1}}\text{I}_{\text{2}}\text{R}$ model are shown in Fig.~\ref{fig:vplot}. For very small $a$ (that is, a very large difference among the strains in terms of the rates) $v(\xi)$ becomes strongly localized at $\xi=-1$, and the minimum value $v_{\text{min}}(\xi)$ approaches a finite value $v_{min} = -4/(3\sqrt{3}) = - 0.7698\ldots$.

The plot of $v(\xi)$  for the  $\text{SI}_{\text{1}}\text{I}_{\text{2}}\text{S}$ model is quite similar, so we do not show it here. The asymptotic minimum value of $v(\xi)$ as $a\to 0$ is equal to $-1/2$ in this case.

\subsection{Fixation probabilities}
Fixation of strain 2 (strain 1) occurs when the effective one-dimensional Markov process, described by Eq.~(\ref{eq:FP-final}), reaches the boundary $\xi = -1$ ($\xi = 1$, respectively).  The probability
$\pi_{-1}(\xi_0)$ that strain 2 fixates, given the initial condition on the CL,
corresponds to the exit at $\xi =-1$ and obeys the ordinary differential equation
\begin{equation}
v(\xi_0) \pi^{\prime}_{-1}(\xi_0) + \frac{1}{2}D(\xi_0) \pi^{\prime \prime}_{-1}(\xi_0) = 0,
\end{equation}
see e.g. \cite{Gardiner}.  In this Section the primes stand for the derivatives with respect to the argument. The boundary conditions are $\pi_{-1}(-1) = 1$ and $\pi_{-1}(1) = 0$.  The solution to this problem is
\begin{equation}
\pi_{-1}(\xi_0) = \frac{\int_{\xi_0}^{1} \nu(x)\,dx}{\int_{-1}^{1} \nu(x)\,dx},
\end{equation}
where
\begin{equation}
\nu(x) = e^{-2\int_{0}^x \frac{v(y)}{D(y)}\,dy}.
\end{equation}
\begin{figure}[ht]
\includegraphics[width=3.2in,clip=]{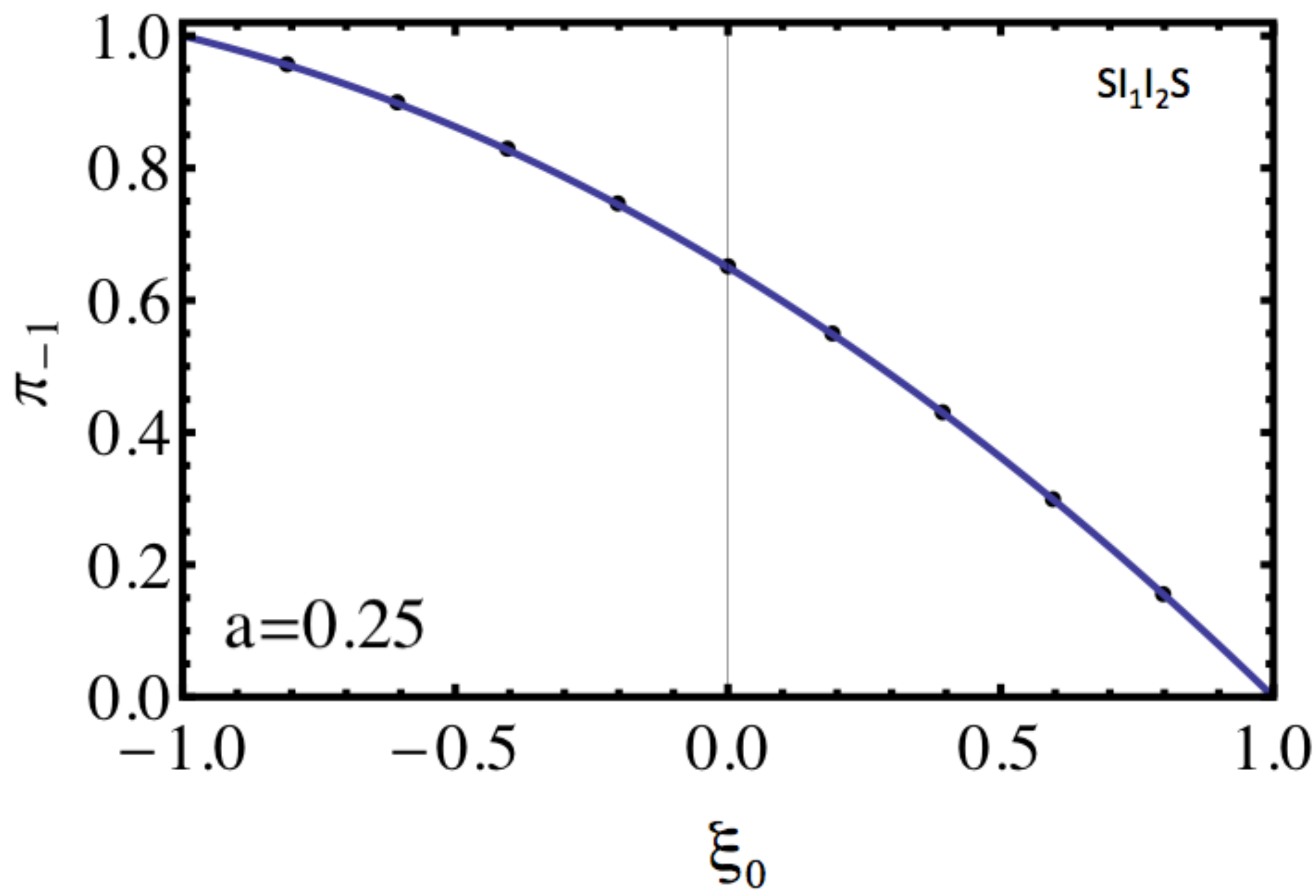}
\includegraphics[width=3.2in,clip=]{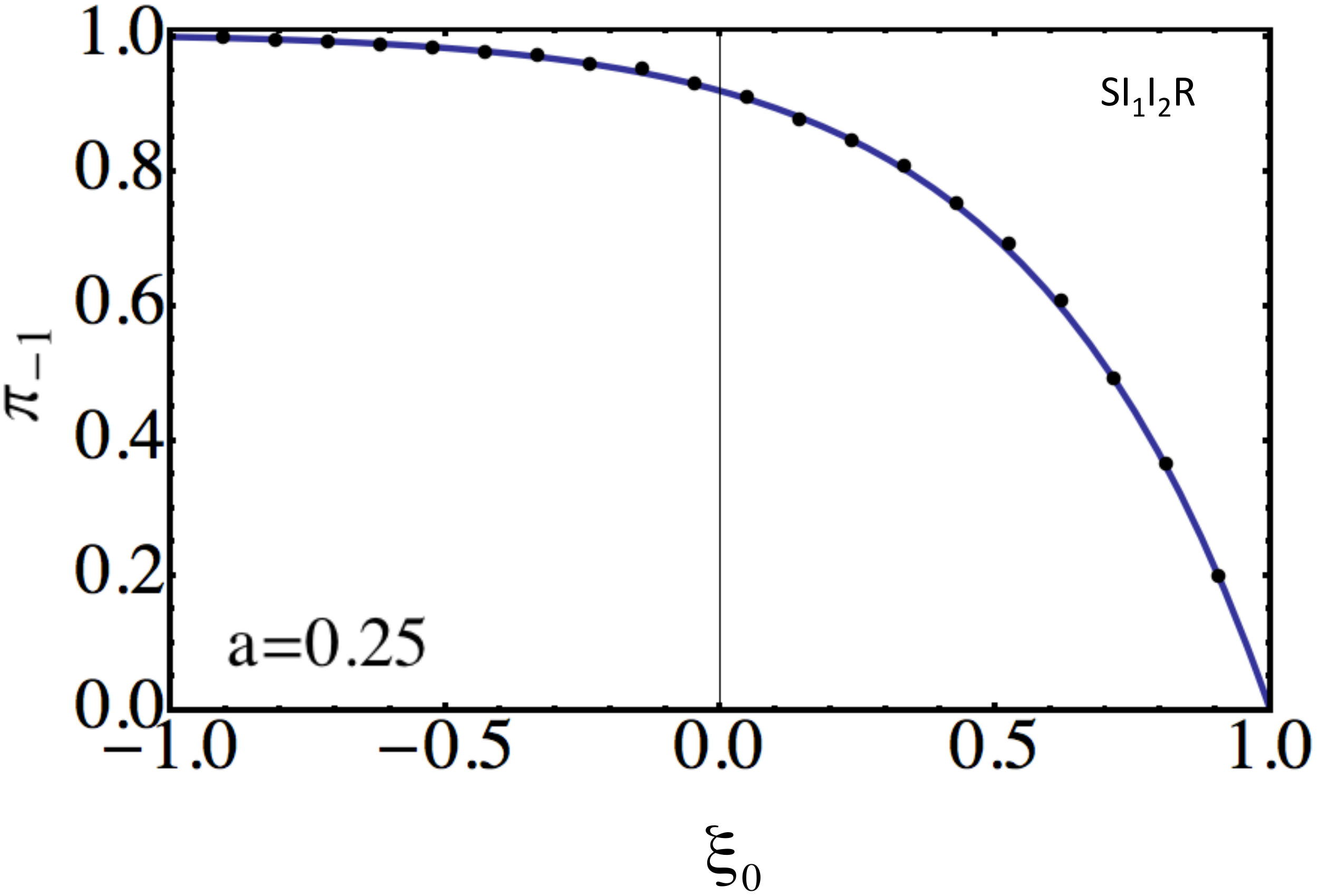}
\caption{(Color online) Fixation probability of the (slower) strain $2$ vs. the initial condition $\xi_0$ on the coexistence line for $a=0.25$,  $\mathcal{R}=4$ and $N=250$. The solid curves were calculated from Eq.~(\ref{eq:pi2d}) for the $\text{SI}_{\text{1}}\text{I}_{\text{2}}\text{S}$ model and from Eq.~(\ref{eq:pi3d}) for the $\text{SI}_{\text{1}}\text{I}_{\text{2}}\text{R}$ model (in the latter case we set $\mu=2$). The dots were obtained in the former case by numerically solving the master equation, Eq.~(\ref{eq:master_master2d}), and in the latter case by averaging over $10^5$ realizations of Monte Carlo simulations.
The curves being convex upward implies a competitive advantage of the slow strain for random initial conditions on the CL.}
\label{prob}
\end{figure}
For the $\text{SI}_{\text{1}}\text{I}_{\text{2}}\text{S}$ model, the solution is
\begin{equation}
\label{eq:pi2d}
\pi_{-1}(\xi_0)=\frac{(1-\xi_0) [(1-a) \xi_0+a+3]}{4 (a+1)},
\end{equation}
in agreement with Refs. \cite{Parsons1,Doering}. For the $\text{SI}_{\text{1}}\text{I}_{\text{2}}\text{R}$ model we obtain
\begin{equation}
\label{eq:pi3d}
\pi_{-1}(\xi_0)= \frac{2 e^{\frac{1}{a}}-e^{\frac{1+ a^2+(1-a^2)\xi _0}{2 a}} \left[1+ a^2+(1-a^2)\xi _0\right]}{2 \left(e^{\frac{1}{a}}-a^2 e^a\right)}.
\end{equation}
Figure~\ref{prob} compares, for a set of parameters, $\pi_{-1}(\xi_0)$ predicted by Eqs.~(\ref{eq:pi2d}) and~(\ref{eq:pi3d})
with $\pi_{-1}(\xi_0)$ obtained by (i) solving the master equation numerically (for the $\text{SI}_{\text{1}}\text{I}_{\text{2}}\text{S}$ model) and by (ii) averaging over $10^5$ realizations of Monte Carlo simulations (for the $\text{SI}_{\text{1}}\text{I}_{\text{2}}\text{R}$ model). The initial conditions, in both cases, where chosen to lie on the CL. For $N=250$ a very good agreement is observed.

The fixation probability of strain 1 is
\begin{equation}
\label{eq:piplus}
\pi_{+1}(\xi_0)=1-\pi_{-1}(\xi_0)\,.
\end{equation}
When the two strains are identical, $a=1$,  we recover the expected results
\begin{eqnarray}
\label{eq:pia1}
\pi_{-1}(\xi_0)=\frac{1}{2}\left(1-\xi_0\right)\,,\;\;\;\;
\pi_{+1}(\xi_0)=\frac{1}{2}\left(1+\xi_0\right)\,,
\end{eqnarray}
for both models.  In this case $\pi_{-1}(0)=\pi_{+1}(0)=1/2$, and the strains are equally competitive.

\subsection{Mean time to fixation}
\begin{figure}[ht]
\includegraphics[width=3.32in, clip=]{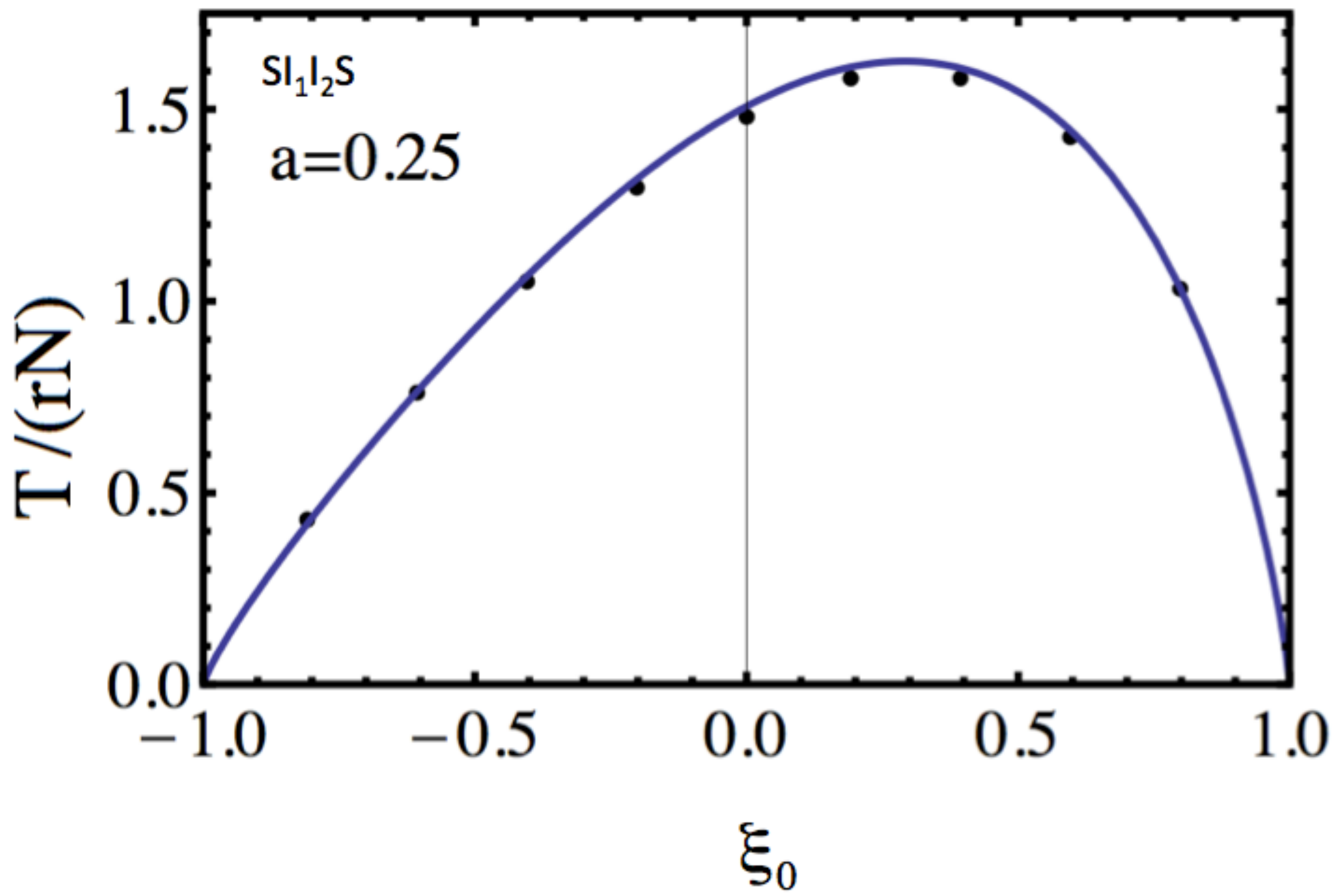}
\includegraphics[width=3.32in, clip=]{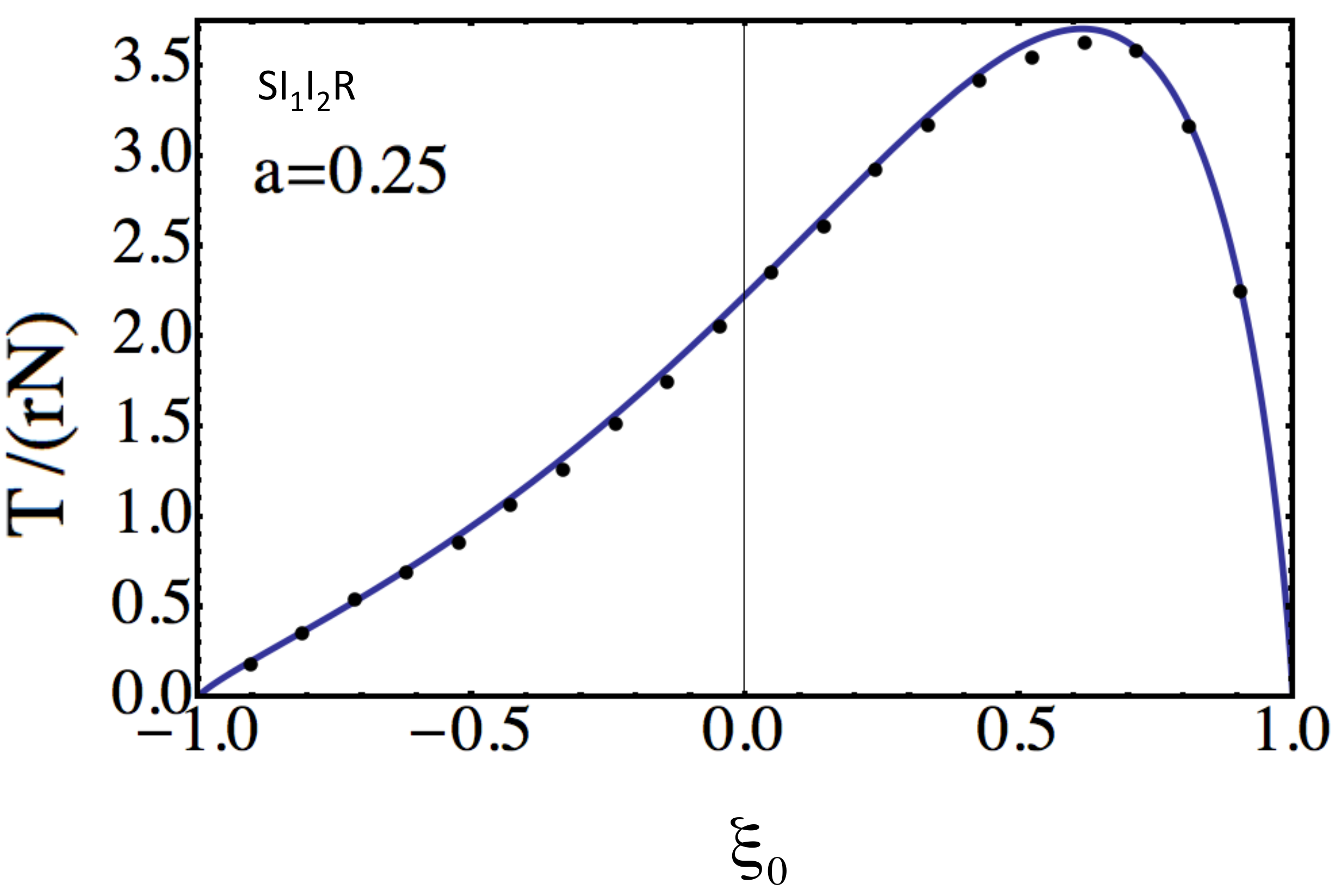}
\caption{(Color online) The rescaled mean time to fixation $T/(r N)$ vs. the initial condition $\xi_0$ on the coexistence line for $a=0.25$, $\mathcal{R}=4$, $N=250$ (and $\mu=2$ for the $\text{SI}_{\text{1}}\text{I}_{\text{2}}\text{R}$ model). The dots were obtained by numerically solving the master equation, Eq.~(\ref{eq:master_master2d}), for the $\text{SI}_{\text{1}}\text{I}_{\text{2}}\text{S}$ model and by averaging over $10^5$ realizations of Monte Carlo simulations of the $\text{SI}_{\text{1}}\text{I}_{\text{2}}\text{R}$ model.}
\label{T}
\end{figure}

The mean time to fixation (MTF) $T(\xi_0)$ obeys the equation
\begin{equation}
\label{eq:eq-for-T}
v(\xi_0) T^{\prime}(\xi_0) + \frac{1}{2}D(\xi_0)T^{\prime \prime}(\xi_0) = -1
\end{equation}
with the boundary conditions $T(-1) = T (1) = 0$, see e.g. Ref. \cite{Gardiner}.  Reintroducing time $t$ as it appears in Eq.~(\ref{eq:mf2d})
or Eq. (\ref{eq:mf}), we can write the solution to this problem as
\begin{eqnarray}
T(\xi_0) &=& rN\int_{-1}^{\xi_0} \nu(x)\left[Q_0- Q(x)\right] \,dx,  \nonumber \\
\nonumber \\
Q(x) &=& \int_{0}^x \frac{2\,dy}{\nu(y)D(y)}, \nonumber \\
Q_0 &=& \frac{\int_{-1}^1 \nu(x) Q(x) \,dx}{\int_{-1}^1 \nu(x)\,dx}. \label{eq:Tgeneral}
\end{eqnarray}
(We remind the reader that the definition of $r$ in the two models differs by a factor $\mu$.)  The integrals in Eq.~(\ref{eq:Tgeneral}) can be evaluated analytically for both models. We discuss some analytic properties of the mean time to fixation in Appendix \ref{appfix}. Figure~\ref{T} compares, for a set of parameters, these analytic results with
numerical results obtained by (i) a numerical solution of the master equation (for the $\text{SI}_{\text{1}}\text{I}_{\text{2}}\text{S}$ model) and (ii) by averaging over $10^5$ realizations of Monte Carlo simulations (for the $\text{SI}_{\text{1}}\text{I}_{\text{2}}\text{R}$ model). The initial conditions, in both cases, lie on the CL. As one can see, for $N=250$ a very good agreement is observed.

\section{Competitive advantage and initial conditions}
\label{sec:CompAdv}
\begin{figure}[h]
\includegraphics[width=2.5in,clip=]{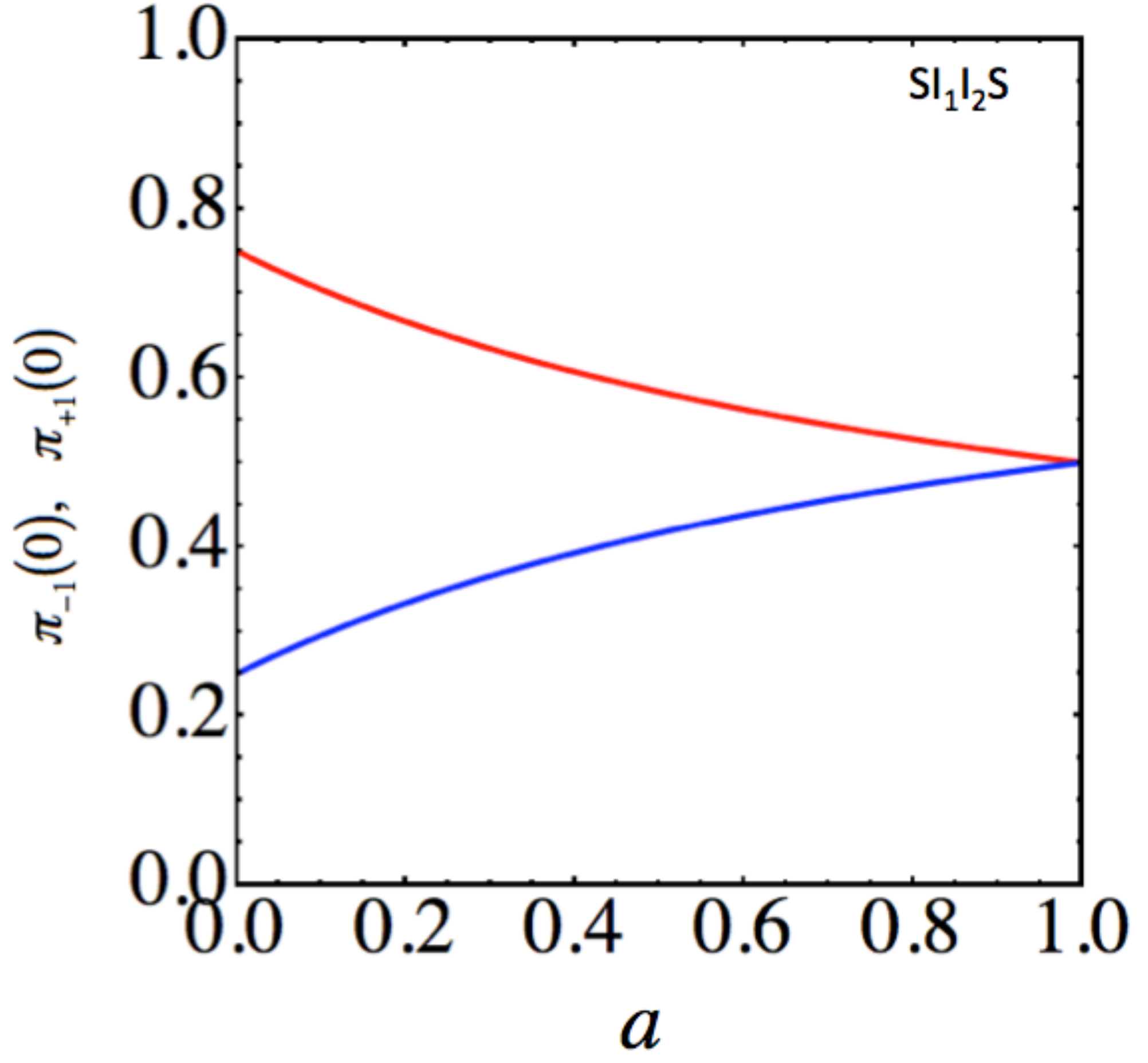}
\includegraphics[width=2.5in,clip=]{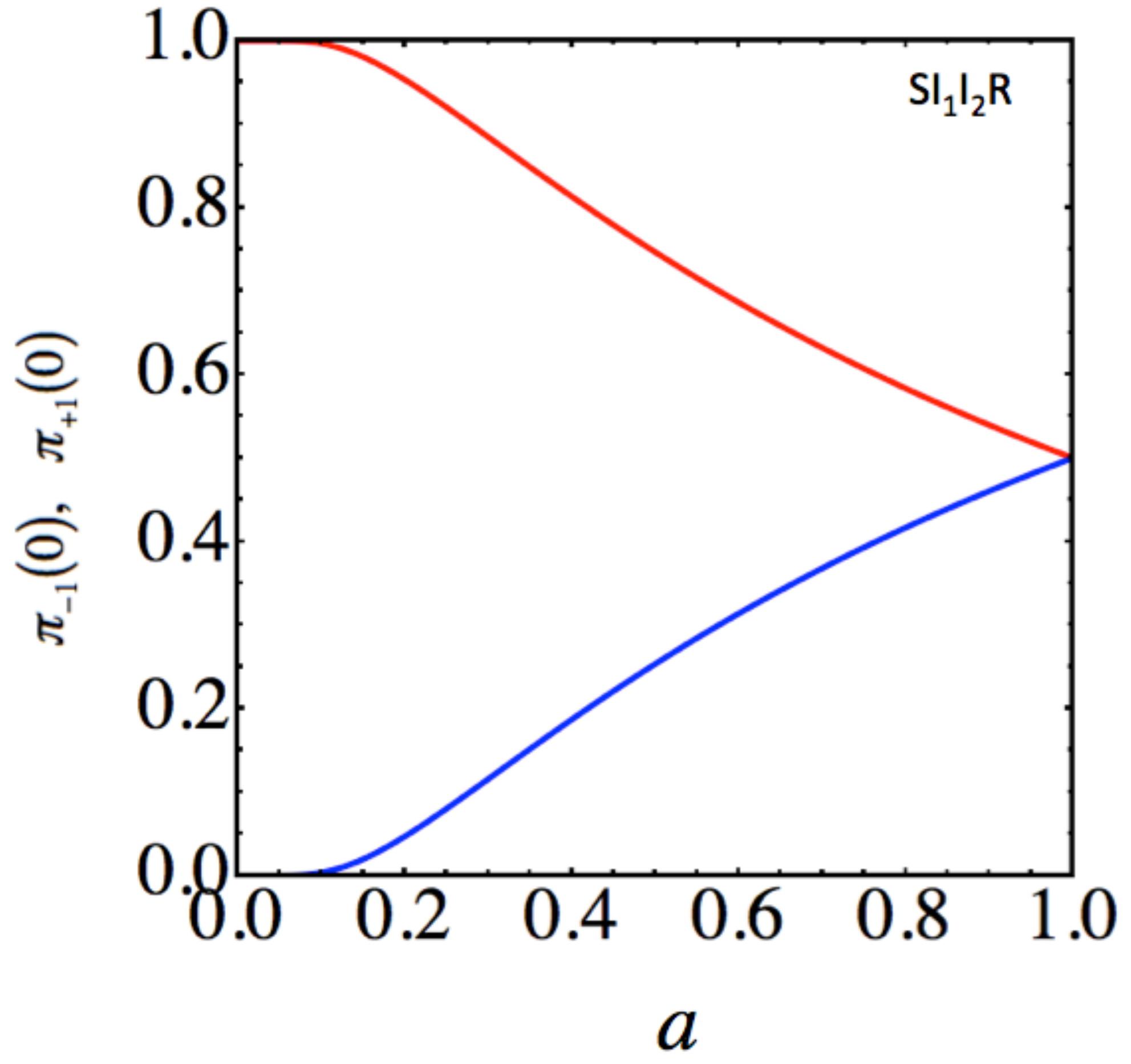}
\caption{(Color online) Extinction probability of each of the two strains in the middle of the coexistence line, $\pi_{-1}(\xi_0 = 0)$ (red) and $\pi_{+1}(\xi_0 = 0)$ (blue) vs. $a$ as a measure of the competitive advantage of the slow strain for initial conditions on the CL, for both models. For $a<1$, $\pi_{-1}(\xi_0 = 0)  > 1/2$.}
\label{comp-a}
\end{figure}
\begin{figure}[h]
\includegraphics[width=2.5in,clip=]{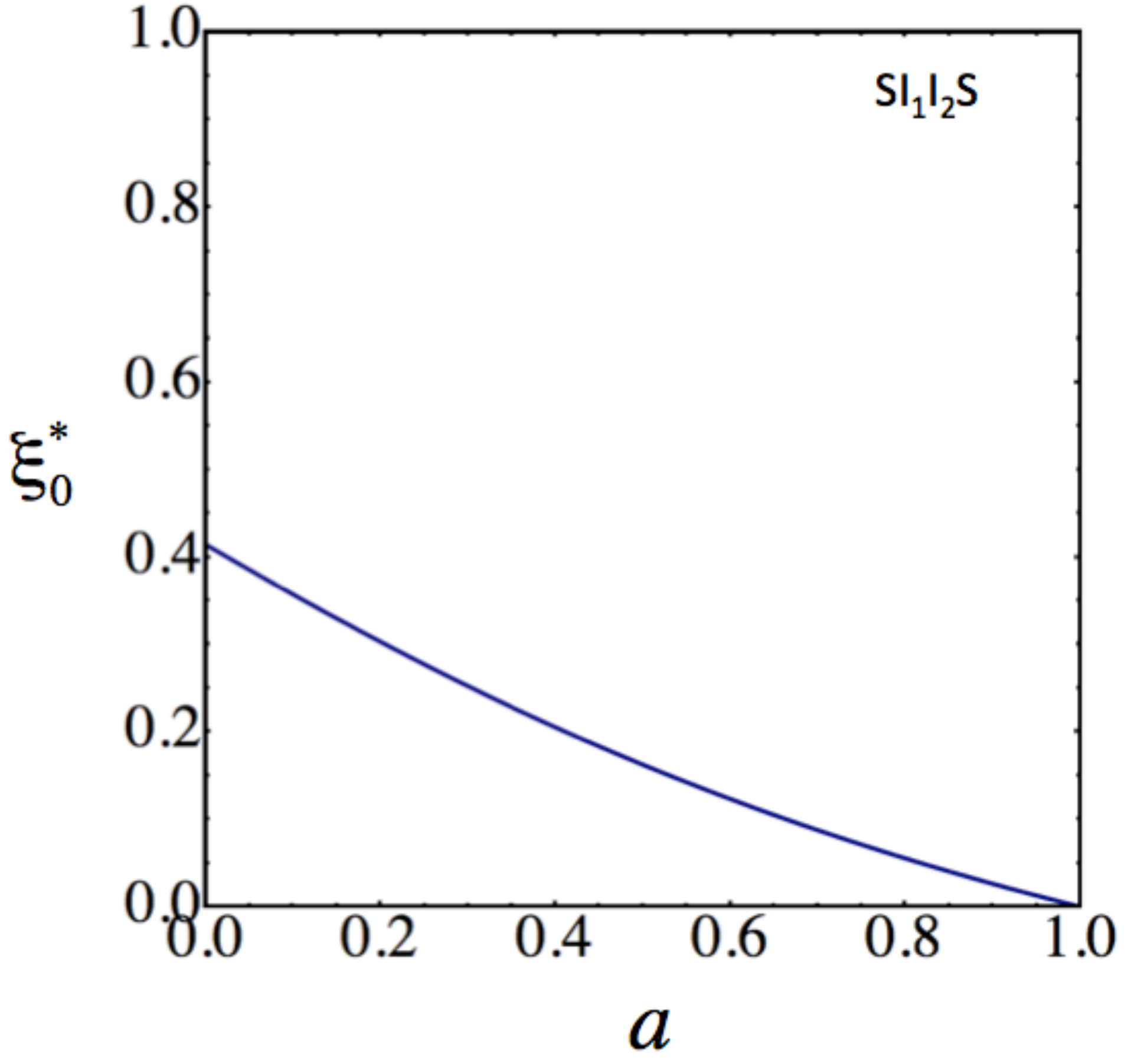}
\includegraphics[width=2.5in,clip=]{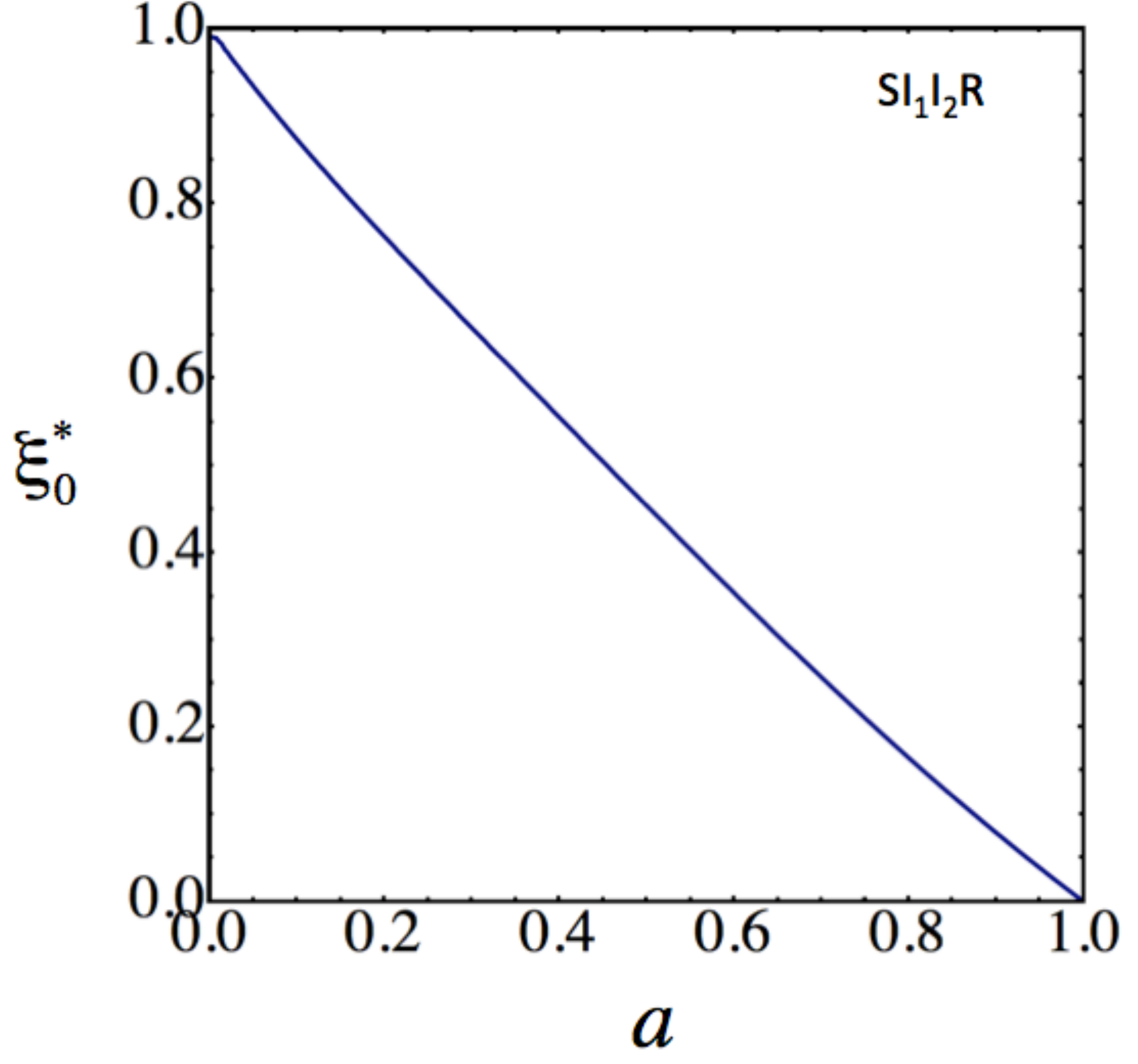}
\caption{(Color online) Neutral position along the coexistence line $\xi_0^*$ vs. $a$ as a measure of competitive advantage of the slow strain with respect to initial conditions on the CL, for both models.}
\label{comp-b}
\end{figure}

One measure of strain competitiveness is $\pi_{\pm 1}(0)$: the fixation probabilities at $\xi_0=0$.  We plot these functions for both models in Fig.~\ref{comp-a}.  One has $\pi_{-1}(0)>1/2$ and $\pi_{+1}(0)<1/2$, that is, the slow strain has an advantage \cite{Parsons1,Parsons2,Doering}.   This effect is especially pronounced in the $\text{SI}_{\text{1}}\text{I}_{\text{2}}\text{R}$ model.

Alternatively, we may find the point $\xi_0^*$ on the CL where $\pi_{\pm 1}(\xi_0^*)=1/2$, so that the fast and slow strains have equal probabilities to fixate from that initial condition on the CL.  For the $\text{SI}_{\text{1}}\text{I}_{\text{2}}\text{S}$ model, Eq.~(\ref{eq:pi2d}) yields
\begin{equation}
\label{xi*}
\xi_0^*=\frac{a+1-\sqrt{2(a^2+1)}}{a-1}\,.
\end{equation}
Interestingly, $\xi_0^*$ reaches a limit distinct from $1$, $\xi_0^* \rightarrow \sqrt{2}-1$, as $a\rightarrow 0$.
The $\text{SI}_{\text{1}}\text{I}_{\text{2}}\text{R}$ model behaves differently at $a\to 0$. Here Eq.~(\ref{eq:pi3d}) yields $\xi_0^* \simeq 1-2a\ln{2}$ at $a\ll 1$, so $\xi_0^* \to 1$ as $a \to 0$.
Using the relations $x = r(1+\xi)/2$ and $y = r(1-\xi)/(2a)$ on the CL of the  $\text{SI}_{\text{1}}\text{I}_{\text{2}}\text{R}$ model, we see that $x^*/r \rightarrow 1$ and $y^*/r \rightarrow \ln{2}$ as $a\rightarrow 0$.  Therefore  $\xi_0^*$ goes to $1$  as $a$ goes to zero  because the slope and the length of the CL grow.
Here, as $a\to 0$, the slow strain fixates for an ever-growing fraction of initial conditions along the CL, see Fig.~\ref{comp-b}. Notice also that $\xi_0\simeq (3/4) (1-a)$ as $a$ approaches 1.

These two closely-related measures of competitiveness assume initial conditions
drawn from the uniform distribution along the CL. To deal with an arbitrary initial condition off the CL,
we define the separating curve
\begin{equation}
\label{limiting}
y_{\text{sep}} = M^*(a) x ^a
\end{equation}
that passes through the point $(x^*, y^*)$  on the CL, see Fig.~\ref{fig:100}.  All initial conditions with $x$ and $y$ above this curve reach the part of the CL where the subsequent stochastic dynamics is more likely to lead to the survival of the slow strain.  All initial conditions that lie below this curve lead to a more likely survival of the fast strain.

Among all possible initial conditions, the more relevant one corresponds to spread of disease when a few infectives of both strains are introduced into a susceptible population. (For our theory to be valid we still assume that the initial number of infected is much larger than 1.) Here the initial conditions are located in the vicinity of the origin in Fig.~\ref{fig:0}, and the
competitive advantage of the strains is determined by the subsequent distribution of the states on the CL following the deterministic evolution.  Assuming a uniform initial distribution of the strains, most of them will be found \emph{under} the separating curve (\ref{limiting}). As a result,  the deterministic evolution brings most of such initial conditions to a point on the CL corresponding to a higher fixation probability of the \emph{fast} strain.   This is a consequence of the (generic) shape of deterministic curves (\ref{phaseplane1}). The fast strain, although in a disadvantage for uniformly distributed initial conditions, becomes advantageous for the more relevant ones.

To come up with a quantitative measure of competitive advantage of the fast strain in this scenario, we consider
\begin{figure}[ht]
\includegraphics[width=2.6in, clip=]{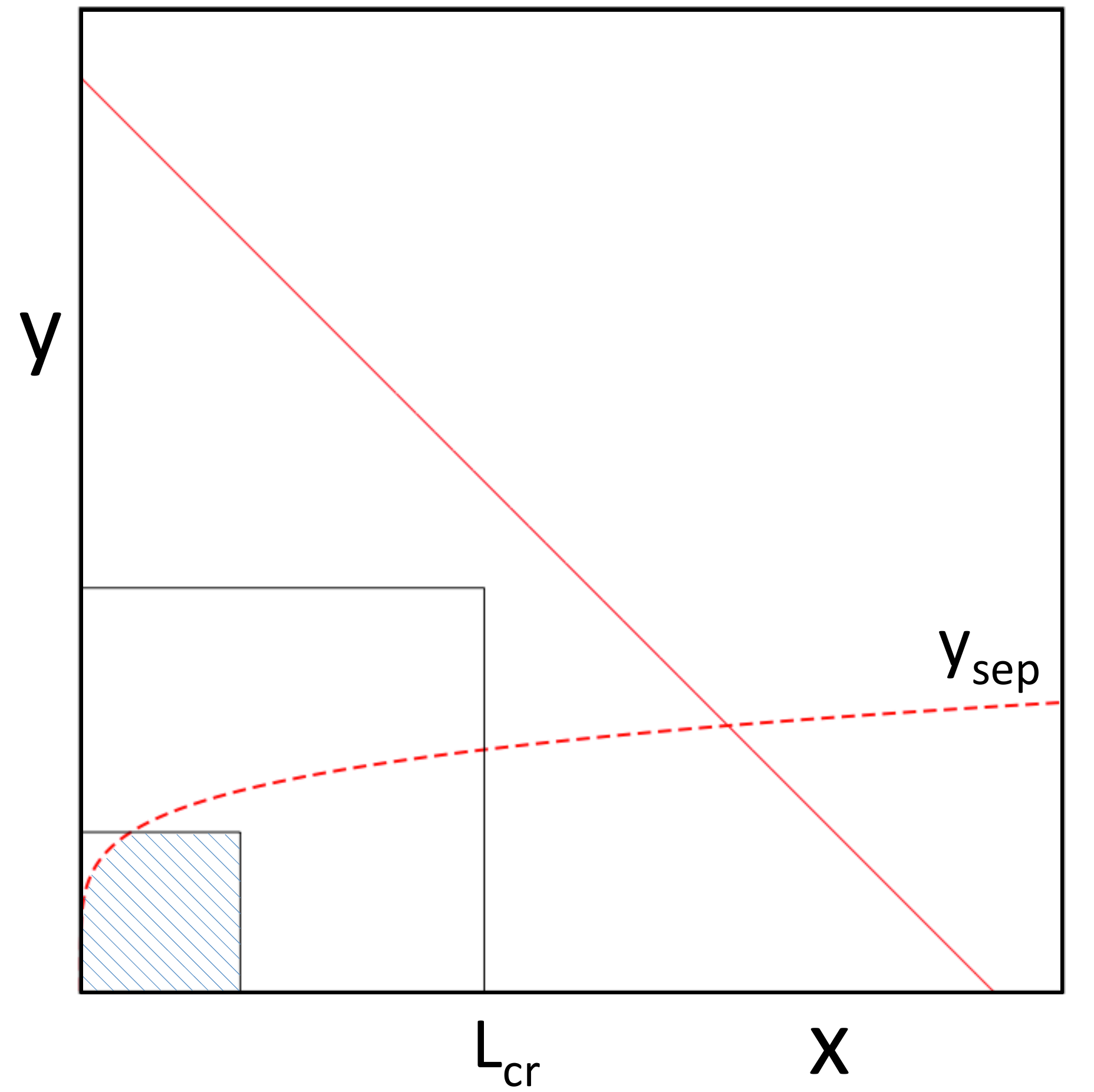}
\caption{(Color online) Competitive advantage of the fast strain in the case when a few infectives of both strains are introduced into a susceptible population. Shown are the CL (solid), the separating curve $y_{\text{sep}}(x)$,
see Eq.~(\ref{limiting})  (dashed), and two different squares of initial conditions.  For $L=L_{\text{cr}}$, half of the square is below $y_{\text{sep}}$. For $L<L_{\text{cr}}$ the fast strain is advantageous as illustrated by the dashed area of the smaller square compared to the empty area of that square.}
\label{SquarePlot}
\end{figure}
\\
a square of size $L$ of initial conditions on the $xy$ plane, as shown in Fig. \ref{SquarePlot}. We can define the competitiveness of the fast strain as the fraction $\mathcal{F}$ of the area of this square
under the separating curve $y_{\text{sep}}(x)$.   Note that $\mathcal{F} \rightarrow 1$ as $L \rightarrow 0$, so the fast strain is always advantageous when the initial number of infected with two strains is sufficiently small. To compute the critical size $L_{cr}$ of the square, or the corresponding critical size of the infected  group $n_{cr}=N L_{cr}$, such that $\mathcal{F} = 1/2$, we first notice that $L_{cr}$ will be greater than the point of intersection of $y_{\text{sep}}(x)$ with the line $y=x$ that happens at $L = (M^*)^{1/(1-a)}$.

For larger $L$  we obtain
\begin{equation}
\mathcal{F} = \frac{\mbox{Area below $y_{\text{sep}}(x)$}}{L^2} = \frac{M^* L^{a-1}}{a+1}.
\end{equation}
Setting $\mathcal{F} = 1/2$ we find
\begin{equation}
\label{eq:Lcr-defn}
L_{cr} = \frac{n_{cr}}{N} = \left(\frac{2M^*}{a+1}\right)^{\frac{1}{1-a}}.
\end{equation}
For $n<n_{cr}$ the fast strain is advantageous.  Expression (\ref{eq:Lcr-defn}) is generic, but the value of $M^*$ is model-dependent.  For the $\text{SI}_{\text{1}}\text{I}_{\text{2}}\text{S}$ model, we can use Eqs.~(\ref{phaseplane}) and (\ref{xi*}) and the relations $x = r(1+\xi)/2$ and $y = r(1-\xi)/2$ on the CL to obtain $M^*(a)$. Inserting it into Eq.~(\ref{eq:Lcr-defn}), we finally obtain
\begin{equation}
\frac{n_{cr}}{N} =   \frac{r}{1-a} \left[\frac{2^a\left(2-\sqrt{2a^2+2}\right)}{(a+1)(\sqrt{2a^2+2}-2a)^a} \right]^{\frac{1}{1-a}}.
\end{equation}
This expression varies monotonically in a narrow range between $2-\sqrt{2} = 0.58578 \ldots $ at $a=0$ and $1/2$ as $a \rightarrow 1$.

In the $\text{SI}_{\text{1}}\text{I}_{\text{2}}\text{R}$ case we use the asymptotics  $\xi_0^*  \simeq 1- 2a \ln{2}$ for small $a$ and $\xi_0^*  \simeq (3/4)(1-a)$ for $a$ close to $1$.  From this we find that $n_{cr}/(Nr)$ varies between $2\ln{2}=1.38629 \ldots $ at $a=0$ and  $1/2$ as $a \rightarrow 1$.  As we see, for both models  the critical size of the introduced infected group for the slow strain to outcompete the fast one is comparable to the total population size. For much smaller infected groups, $1\ll n \ll n_{cr}$, the fast strain is much more competitive. In this regime, the probability of fixation of the fast strain behaves like
\begin{equation}
\mathcal{F} \simeq 1-(n/N)^{1/a},
\end{equation}
which very rapidly approaches $1$ as $n/N$ becomes small.

We briefly mention another type of initial condition. The system may initially have only one strain, so that it is positioned at the end of the CL. Then a mutation, or a one-time importation occurs that introduces a minority strain and causes a small deviation of the system into the interior of the $xy$ plane.  The subsequent stochastic dynamics will most likely bring the system back to the original end of the CL. There is a small but finite probability, however, that the system will wander along the CL and switch to the minority strain. This probability scales as $n/N$, where $n$ is the number of individuals with a new strain.

Finally, to compute the survival probability of the fast strain for an arbitrary initial condition $(x_0,y_0)$, one needs to solve numerically for $\xi_0$ from
\begin{eqnarray}
\mbox{$\text{SI}_{\text{1}}\text{I}_{\text{2}}\text{S}$:~~~~} \frac{y_0}{x_0^a} &=& \frac{r^{1-a}(1-\xi_0)}{2^{1-a}(1+\xi_0)^a},\\
\mbox{$\text{SI}_{\text{1}}\text{I}_{\text{2}}\text{R}$:~~~~} \frac{y_0}{x_0^a} &=& \frac{r^{1-a}(1-\xi_0)}{2^{1-a} a (1+\xi_0)^a},
\end{eqnarray}
and substitute this value into expression $\pi_{+1}(\xi_0) = 1-\pi_{-1}(\xi_0)$ from Eqs~(\ref{eq:pi2d}) or (\ref{eq:pi3d}).

\section{Discussion}
\label{sec:Discussion}
It is well known by now that shot noise can cause qualitative, and sometimes dramatic, changes in the system's behavior compared with predictions of a deterministic theory.  In the two-strain variants of the SIS and SIR models that we have considered, the competition for resources is neutral if the noise is neglected. When it is taken into account, it determines the outcome of the competition: extinction of one strain and fixation of the other. Because of the noise, one of the strains turns out to have an advantage in fixation, depending on the type of initial conditions. The slow strain has a competitive advantage for uniformly distributed initial conditions. It is the fast strain, however, that is more likely to win in the practically important situation when a few infectives of both strains are introduced into a susceptible population. In fact, the fast strain remains advantageous for relatively large numbers of infectives of both strains that scales as the population size. These results are generic and expected to be valid for many additional models of quasi-neutral competition in epidemiology, population biology and population genetics.

At a technical level, we obtained these and other results by developing a novel perturbation method that employs the smallness of the parameter 1/$N$ and reduces, in a systematic way, a multi-dimensional master equation to an effective one-dimensional Fokker-Planck equation along the coexistence line (CL) of the quasi-neutral model. The method also describes the whole multi-dimensional probability distribution of the system in the vicinity of the CL.  We note that the method is similar in spirit to the Born-Oppenheimer approximation of quantum mechanics \cite{Born}. We expect it to be applicable to a whole class of quasi-neutral competition models in different fields of science.

From a broader perspective, quasi-neutral competition is an instance of a general scenario where a weak shot noise has a large accumulated effect on nonlinear systems when it acts in directions with zero eigenvalues. The noise causes diffusion and (positive or negative) drift in these directions.
A classic example of this scenario is phase diffusion  \cite{phasediffusion,phasedrift} and phase drift \cite{phasedrift} of noisy limit cycles.
Another example is the Lotka-Volterra predator-prey model, where shot noise causes slow diffusion and drift across neutral cycles of the deterministic theory,
and ultimately causes extinction or proliferation of the species \cite{PK}. A spatially explicit example is
the shot-noise-induced velocity fluctuations of population invasion fronts which include both diffusion \cite{MSK,KhM} and a systematic drift \cite{KhM} of the front position compared to predictions from deterministic theory.

A future work can explore quasi-neutral competition in spatial systems, as envisioned in Ref. \cite{KS}.

\section*{Acknowledgments}

We are grateful to Charles R. Doering and Leonard M. Sander for attracting our interest in quasi-neutral competition,
and to Michael C. Cross and Mark I. Dykman for a discussion of phase diffusion and drift in noisy limit cycles.
This research was supported in part by Grant No. 2012145 from the United States-Israel Binational Science Foundation (BSF), and in part by the Science \& Technology Directorate, Department of Homeland Security via interagency agreement no.\ HSHQDC-10-X-00138.
BM acknowledges the hospitality of the Michigan Center for Theoretical Physics where this project was started.

\onecolumngrid
\appendix

\section{Linear stability of the CL of the $\text{SI}_{\text{1}}\text{I}_{\text{2}}\text{R}$ model}
\label{sec:MFAppendix}
In the variables $(Z,Y^{\prime},Z^{\prime})$, see Eq.~(\ref{eq:newvarnoN}), the deterministic equations have the following form:
\begin{eqnarray}
\label{eq:mfXYZ}
\dot{X} &=& \frac{a\mathcal{R}}{(a^2 + 1)}\left[(a+1)XZ^{\prime} + (a-1)Y^{\prime}Z^{\prime} + r(a-1)Z^{\prime}\right], \nonumber\\
\dot{Y^{\prime}} &=& \frac{\mathcal{R}}{(a^2 + 1)}\left[a(a-1) XZ^{\prime} +  (1+a^3)Y^{\prime}Z^{\prime} + r(1+a^3)Z^{\prime} \right], \nonumber \\
\dot{Z^{\prime}} &=& - Y^{\prime} - \mu \mathcal{R} Z^{\prime} - \mathcal{R} Y^{\prime}Z^{\prime}.
\end{eqnarray}
We can linearize these equations around any point $(X, 0, 0)$.  The resulting linear stability matrix is
\begin{displaymath}
\left(\begin{array}{ccc}
0 & 0 & M_{XZ^{\prime}} \\
0 & 0 & M_{YZ^{\prime}}\\
0 & -1 & -\mu \mathcal{R}
\end{array}\right),
\end{displaymath}
where
\begin{eqnarray*}
M_{XZ^{\prime}} &=& \frac{a\mathcal{R}}{1+a^2} \left[(a-1)r + (a+1) X\right],\\
M_{YZ^{\prime}} &=& \frac{\mathcal{R}}{1+a^2} \left[(a^3+1)r + a(a-1)X\right].
\end{eqnarray*}
One of the three eigenvectors is obviously $(1,0,0)$, with the eigenvalue $0$.  The other two eigenvalues obey
\begin{equation}
\label{eq:evals}
\lambda_{\pm}= -\frac{\mu \mathcal{R}}{2} \pm \sqrt{\left(\frac{\mu \mathcal{R}}{2}\right)^2 -  4 M_{YZ^{\prime}}}
\end{equation}
As $M_{YZ^{\prime}}>0$, each of these $\lambda_{\pm}$ always has a negative real part.  The imaginary part may or may not be zero depending on the parameters and on the coordinate $X$ along the CL.
In general, the eigenvalues $\lambda_{\pm}$ are unrelated to inverse widths of the bi-Gaussian $\Lambda_{\pm}$ given by Eq.~(\ref{eq:Lambdapm}).

The eigenvectors, corresponding to $\lambda_{\pm}$, are the following:
\begin{equation}
\label{eq:CL-eigenvectors}
\vec{v} = \frac{1}{\nu} \left(\begin{array}{c} \frac{M_{XZ^{\prime}}}{M_{YZ^{\prime}}} \\ \\ 1 \\ \\ \frac{\lambda}{M_{YZ^{\prime}}} \end{array}\right),
\end{equation}
where the normalization factor $\nu$ is chosen so that the projection of $\vec{v}$ onto the $(Y^{\prime},Z^{\prime})$ plane is normalized:
\begin{equation}
\nu = \sqrt{1 + \left(\frac{\lambda}{M_{YZ^{\prime}}}\right)^2}.
\end{equation}
The transformation from these eigencoordinates  $(s,p,m)$ (for ``slow'', ``plus'', and ``minus'', respectively) to the orthogonal coordinates $(X,Y^{\prime},Z^{\prime})$ is accomplished via
\begin{equation}
\label{eq:eigen-transformation-real}
\left(\begin{array}{c} X \\  \\ Y^{\prime} \\ \\ Z^{\prime} \end{array} \right) =
\left(
\begin{array}{ccccc}
1 &  &  \frac{M_{XZ^{\prime}}}{\nu_+M_{Y^{\prime}Z^{\prime}}} & &  \frac{M_{XZ^{\prime}}}{\nu_-M_{Y^{\prime}Z^{\prime}}}  \\
\\
0 & & \frac{1}{\nu_+} & & \frac{1}{\nu_-} \\
\\
0 & & \frac{\lambda_+}{\nu_+M_{Y^{\prime}Z^{\prime}}} & & \frac{\lambda_-}{\nu_-M_{Y^{\prime}Z^{\prime}}} \\
\end{array}\right) \left(\begin{array}{c} s \\  \\ p \\ \\ m \end{array}\right),
\end{equation}
where $+$ and $-$ label the corresponding eigenvalue.  If the eigenvalues are complex, the attracting manifold of the CL is spanned by the real and imaginary parts of the (complex-conjugate pair of) eigenvectors.  In that case one should replace the ``$+$'' and ``$-$'' subscripts by Re and Im, respectively. The eigenvectors $\vec{V}_{\pm}$ do not coincide with the (normalized to unity) $Y^{\prime}$$Z^{\prime}$ component of the eigenvectors $\vec{v}_{pm}$ from Eq.~(\ref{eq:CL-eigenvectors}).

\section{Operators}
\label{sec:Operators_appendix}
Here we present the explicit forms of the operators $\Op L^{(n)}$, $n=0,1,2$ that appear in our calculations.
\subsection{Operators in the $\text{SI}_{\text{1}}\text{I}_{\text{2}}\text{S}$ model}
\label{sec:Operators_appendix_2d}
\begin{eqnarray}
\Op L^{(0)} \rho&=&\frac{h(X)}{2}\left[\frac{\partial^2 \rho}{\partial Y^2}+\mathcal{R}\frac{\partial}{\partial Y}\left(Y\rho \right)\right], \label{zero}
\end{eqnarray}
\begin{eqnarray}
\label{eq:L12ddefined}
\Op L^{(1)} \rho&=&\frac{(1+a)\mathcal{R}}{2} \frac{\partial}{\partial Y}\left(Y^2 \rho \right)+
\frac{\mathcal{R} Y}{2}\frac{\partial}{\partial X} \left[ g(X)\rho \right] \nonumber \\
&+& \frac{\partial^2}{\partial X \partial Y}\left[g(X) \rho\right] +\frac{1}{2}\frac{\partial^2}{\partial Y^2} \left[Y H(X)\rho \right], \label{first}
\end{eqnarray}
\begin{eqnarray}
\Op L^{(2)} \rho&=&\frac{(1-a)\mathcal{R} Y^2}{2}\frac{\partial \rho}{\partial X}+
\frac{1}{2} \frac{\partial^2}{\partial X^2}\left[h(X)\rho\right]+ \frac{\partial}{\partial_Y}\left\{...\right\}\,, \label{second}
\end{eqnarray}
\begin{eqnarray}
H(X)&=&(1+a)\left(1-\frac{\mathcal{R}r}{2}\right)-\frac{\mathcal{R} X }{2}(1-a)\,, \label{H}
\end{eqnarray}
where $h(X)$ and $g(X)$ are defined in Eqs.~(\ref{eq:nextorder10}) and (\ref{eq:Fsol}), respectively.
The non-specified term in Eq.~(\ref{second}) does not contribute to the integral in Eq.~(\ref{eq:secondb}), since it is a total derivative.  For the same reason only the second term in Eq.~(\ref{first}) contributes.

\subsection{Operators in the $\text{SI}_{\text{1}}\text{I}_{\text{2}}\text{R}$ model}
\label{sec:Operators_appendix_3d}
\begin{equation}
\label{eq:Operator_L0}
\hat{L}^{(0)}\rho = \mathcal{R} d_0(X)  Z\partial_Y \rho + \partial_Z \left[(\mu \mathcal{R} Z + Y)\rho\right] + c_0(X) \partial^2_Y\rho + \mu \partial^2_Z\rho + d_0(X) \partial^2_{YZ}\rho,
\end{equation}
where
\begin{eqnarray}
\label{eq:c-d-coeffs}
c_0 (X)&=&  \frac{1}{1+a^2}\left[(1+a^4)r + a(a^2-1)X\right], \nonumber \\
d_0(X)&=&-\frac{1}{1+a^2}\left[(1+a^3)r + a(a-1)X\right]. 
\end{eqnarray}
In this operator only $Y$ and $Z$ are independent variables, while $X$ is ``frozen".  Due to introduction of the small parameter $\varepsilon$ upon rescaling the transverse variables, only the linearized terms of the full deterministic equations in variables $(X,Y^{\prime},Z^{\prime})$ contribute to the drift terms in $\Op L^{(0)}$.  By the same token, only a subset of the diffusion coefficients of the full Fokker-Planck operator in variables $(X,Y^{\prime},Z^{\prime})$ is present in $\Op L^{(0)}$. Further,
\begin{eqnarray}
\label{eq:L1}
\hat{L}^{(1)}\rho &=& \frac{\mathcal{R}a}{1+a^2} \partial_X \left\{\left[(1-a)r - (1+a)X \right]Z\rho\right\} - \mathcal{R}\frac{1+a^3}{1+a^2}\partial_Y \left(YZ\rho\right) + \mathcal{R}\partial_Z \left(YZ\rho\right)  \nonumber \\
&+& \frac{1}{1+a^2}\partial^2_Y \left\{\left[\mathcal{R}\frac{XZ}{2}a(a^2-1) + \mathcal{R}\frac{rZ}{2}(a^4 + 1)+(1+a^4)Y\right]\rho\right\}  +\frac{1}{2}\partial^2_Z \left[\left(\mathcal{R}\mu Z + Y\right)\rho\right] \nonumber \\
&+& \frac{2a}{1+a^2} \partial^2_{XY}\left\{\left[2aX + r(a^2-1)\right]\rho\right\} + \frac{1}{1+a^2} \partial^2_{XZ} \left(\left\{a[r(1-a) - (1+a)X] - (1+a^3)Y\right\}\rho\right) \nonumber  \\
&+& \frac{\mathcal{R}}{1+a^2}\partial^2_{YZ}\left\{\left[a(1-a)XZ - rZ (1+a^3)\right]\rho \right\},
\end{eqnarray}
\begin{eqnarray}
\label{eq:L2}
\hat{L}^{(2)}\rho &=& \mathcal{R}\frac{a(1-a)}{1+a^2}\partial_X\left(YZ\rho\right) + \frac{a}{1+a^2}\partial^2_X\left\{\left[2ar + (1-a^2)X\right]\rho\right\} + \frac{\mathcal{R}}{2}\frac{1+a^4}{1+a^2}\partial^2_Y\left(YZ\rho\right) + \frac{\mathcal{R}}{2} \partial^2_Z \left(YZ\rho\right) \nonumber \\
&+& \frac{1}{1+a^2}\partial^2_{XY}\left(\left\{\mathcal{R}a[2aX - (1-a^2)r]Z + (1+a^2)Y\right\}\rho \right) + \frac{a}{1+a^2}\partial^2_{XZ}\left(\left\{(1-a)Y + \mathcal{R}\left[(1-a)r - (1+a)X\right]Z\right\}\rho\right) \nonumber \\
&-& \mathcal{R}\frac{1+a^3}{1+a^2}\partial^2_{YZ}\left(YZ\rho\right).
\end{eqnarray}
Once again, most terms are total derivatives with respect to integration in $Y$ and $Z$ variables in Eq.~(\ref{eq:solvability_condition}).  Only the first term in Eq.~(\ref{eq:L1}) and only the first two terms in Eq.~(\ref{eq:L2}) will contribute, giving rise to Eq.~(\ref{eq:F3d}) and Eq.~(\ref{eq:second_term}) respectively.

\section{A general framework for the derivation of the 1D Fokker-Planck equation}
\label{bypass}
\subsection{General formulation}
\label{GenForm}
Our starting point in this Appendix is the equation
\begin{equation}
\label{geneq:exp2d}
\partial_t \rho(X,Y,t) = \left(\Op L^{(0)}+\varepsilon \Op L^{(1)}+ \varepsilon ^2 \Op L^{(2)}\right)\rho (X,Y,t),\;\;\;\varepsilon=1/\sqrt{N} \ll 1\,.
\end{equation}
This equation generalizes the Fokker-Planck equation, Eq.~(\ref{eq:exp2d}), to include a whole class of quasi-neutral competition models, where  $X$  stands for the slow variable (along the CL), and $Y$  stands for the set of \emph{all} fast variables (perpendicular to the CL), unless they are identified explicitly as $Y_i$.

Being interested in the solution of Eq.~(\ref{geneq:exp2d}) that develops on a slow time scale of ${\mathcal O}(\varepsilon^{-2})= \mathcal O(N)$, we make the ansatz
\begin{equation}
\label{ansatz1}
\rho(X,Y,t) = \rho^{(0)}(X, Y, \varepsilon^2 t)+\varepsilon \rho^{(1)}(X,Y,\varepsilon^2 t) + \varepsilon^2 \rho^{(2)}(X,Y,\varepsilon^2 t)+ \ldots .
\end{equation}
Plugging it into Eq.~(\ref{geneq:exp2d}) we obtain
\begin{equation}
\label{geneq:zero}
\Op L^{(0)} \rho^{(0)}=0
\end{equation}
in the zeroth order of $\varepsilon$,
\begin{equation}
\label{geneq:first}
\Op L^{(0)}\rho^{(1)}=-\Op L^{(1)}\rho^{(0)}
\end{equation}
in the first order of $\varepsilon$, and
\begin{equation}
\label{geneq:second}
\Op L^{(0)}\rho^{(2)}=\partial_{\tau} \rho^{(0)} - \Op L^{(1)}\rho^{(1)}-\Op L^{(2)}\rho^{(0)}
\end{equation}
in the second order of $\varepsilon$.  Here $\tau=\varepsilon^2 t=t/N$ is the slow time. We shall assume the following structure of the operator $\Op L^{(0)}$:
\begin{eqnarray}
\label{geneq:adjoint_structure2}
&&\Op L^{(0)}=\sum_{ij} \Psi_{ij}(X)\partial_{Y_i} Y_j + \Phi_{ij}(X) \partial^2_{Y_i  Y_j} .
\end{eqnarray}
This form is a general consequence of the balance of powers of $\varepsilon$ in the derivation of Eq.(\ref{geneq:exp2d}).  It implies that the solution to Eq.~(\ref{geneq:zero})
can be written as a Gaussian distribution near the CL:
\begin{eqnarray}
\label{geneq:zerosol}
&&\rho^{(0)}(X,Y, \tau)= f(X,\tau) \,{\cal{N}}(X)\, e^{-\frac{1}{2}\sum_{ij}C_{ij}(X)Y_i Y_j}, \\
&&{\cal{N}}(X)^{-1}=\int dY e^{-\frac{1}{2}\sum_{ij}C_{ij}(X)Y_i Y_j}
\end{eqnarray}
where $f(X,\tau)$ is an arbitrary function. The function $\rho^{(0)}(X,Y,\tau)$, with yet unknown $f(X,\tau)$, is a sharp Gaussian of width $\sim N^{-1/2}$ with respect to $Y^{\prime}$.  Computing the matrix $C_{ij}(X)$ from Eq.~(\ref{geneq:zero})  reduces to the following  linear algebraic problem:
\begin{eqnarray}
\label{geneq:trace}
&&C^{-1}\Psi+\Psi C^{-1}=2\Phi,
\end{eqnarray}
where matrices  $\Phi_{ij}$ and $\Psi_{ij}$ are defined by (\ref{geneq:adjoint_structure2}).

The slow temporal dynamics of $\rho^{(0)}$, i.e. the function $f(X,\tau)$ is described by Eq.~(\ref{geneq:second}).  Integrating the latter equation over the fast variables we obtain
\begin{equation}
\label{geneq:secondb}
 \int_{-\infty}^{\infty} \partial_{\tau}\rho^{(0)} \, dY=   \partial_{\tau}f = \int_{-\infty}^{\infty} \left(\Op L^{(1)}\rho^{(1)}+\Op L^{(2)}\rho^{(0)}\right)\,dY.
\end{equation}
The  integration of the second term on the right hand side of Eq.~(\ref{geneq:secondb}) reduces to the computation of second  moments $\left\langle {Y_i Y_j}\right\rangle$ of the Gaussian distribution (\ref{geneq:zerosol}).

As explained in the main text, a straightforward way to evaluate the integral of the first term on the r.h.s. of Eq.~(\ref{geneq:secondb}) would be to first solve Eq.~(\ref{geneq:first}) for $\rho^{(1)}$. However, this can be hard to do in multi-dimensional problems. The bypass that we now present enables one to avoid solving for $\rho^{(1)}$, by exchanging it for $\rho^{(0)}$.

We assume the following structure of the operator $\Op L^{(1)}$:
\begin{equation}
\label{geneq:adjoint_structure}
\Op L^{(1)}=\partial_X\sum_i \Xi_i(X)Y_i+\text{total derivatives with respect to}\; Y.
\end{equation}
This property holds both for the $\text{SI}_{\text{1}}\text{I}_{\text{2}}\text{S}$ model, and for the $\text{SI}_{\text{1}}\text{I}_{\text{2}}\text{R}$ model with population turnover. In general, it can be justified as follows. The $\partial^2/\partial X^2$ derivative is absent in $\Op L^{(1)}$ since it comes from the diffusion term in the full Fokker-Planck equation which scales as $\varepsilon^2$ [see, for example Eq.~(\ref{eq:fp2d})]. The exact form of the total  derivatives  with respect to  Y terms does not matter since they fall upon the integration in Eq.~(\ref{geneq:secondb}). The first term in Eq.~(\ref{geneq:adjoint_structure}) comes from the drift terms of the full Fokker-Planck equation; to be of order $\varepsilon$ it must be linear in $Y$.

Next, let us define $\Op L^{(n) \dagger}_Y$,
$n=0,1,2$,  to be the linear differential operators adjoint to $\Op L^{(n)}$ with respect to integration over the fast variables $Y$, i.e.
\begin{equation}
\label{geneq:adjoint}
\int_{-\infty}^{\infty}f_1(X,Y)\,\Op L^{(n)}f_2(X,Y) d Y =\int_{-\infty}^{\infty}\Op L^{(n)\dagger}_Yf_1(X,Y)\, f_2(X,Y)  d Y.
\end{equation}
We emphasize that, by definition, the operator $\Op L^{(n)\dagger}_Y$ in Eq.~(\ref{geneq:adjoint}) acts on the fast variables $Y$ of $f_1(X,Y)$ and on the slow variable $X$ of $f_2(X,Y)$.
Therefore, the order in which the functions $f_1$ and $f_2$ appear in the second line of Eq.~(\ref{geneq:adjoint}) is important. It is also worth noticing that
the operator $\Op L^{(0)\dagger}_Y$ involves differentiation only with respect to the fast variables $Y$. Let us define the function $F(X,Y)$ as a forced solution of the inhomogeneous linear partial differential equation
\begin{eqnarray}
\label{geneq:F}
\Op L^{(0)\dagger}_Y  F(X,Y)=\sum_i \Xi_i(X)Y_i.
\end{eqnarray}
With these definitions we obtain
\begin{eqnarray}
\label{geneq:adjoint_structureb}
&&\int_{-\infty}^{\infty}\Op L^{(1)}\rho^{(1)}(X,Y) dY\nonumber \\
&&\stackrel{*}{=}\partial_X \int_{-\infty}^{\infty} \left[\Op L^{(0)\dagger}_Y F(X,Y)\right] \rho^{(1)}(X,Y) dY  \nonumber \\
&&=\partial_X \int_{-\infty}^{\infty} F(X,Y)  \left[\Op L^{(0)}\rho^{(1)}(X,Y)\right] dY  \nonumber \\
&&\stackrel{**}{=}-\partial_X  \int_{-\infty}^{\infty} F(X,Y) \left[\Op L^{(1)}\rho^{(0)}(X,Y)\right] dY \nonumber \\
&&=-\partial_X \int_{-\infty}^{\infty} \left[\Op L^{(1)\dagger}_Y F(X,Y)\right] \rho^{(0)}(X,Y) dY\,,
\end{eqnarray}
where  the starred equality follows from Eqs.~(\ref{geneq:adjoint_structure}), (\ref{geneq:adjoint_structure2}) and (\ref{geneq:F}), and the double-starred equality follows from Eq.~(\ref{geneq:first}). To remind the reader,  $\Op L^{(1)\dagger}_Y$ in Eq.~(\ref{geneq:adjoint_structureb}) acts only on the $Y$ coordinates of $F$ and only on the $X$ coordinate of $\rho^{(0)}$.

Once Eq.~(\ref{geneq:F}) for the function $F(X,Y)$ is solved, we evaluate the integral in Eq.~(\ref{geneq:adjoint_structureb}) and complete the derivation of the effective one-dimensional Fokker-Planck equation,  Eq.~(\ref{geneq:secondb}).  The great advantage of this formalism is that Eq.~(\ref{geneq:F}) for the function $F(X,Y)$ is generally much easier to solve than Eq.~(\ref{geneq:first}) for $\rho^{(1)}$.
Indeed, it follows from Eqs.~(\ref{geneq:adjoint_structure2}) and (\ref{geneq:F})  that
\begin{eqnarray}
&&F(X,Y)=\sum_i a_i(X) Y_i,\label{geneq:F_sol} \\
&&\sum_i a_i(X) \Psi_{ij}(X) =-\Xi_j(X), \label{geneq:a_sol}
\end{eqnarray}
where $\Psi_{ij}$ is  defined by (\ref{geneq:adjoint_structure2}).

We see that the derivation of the $1$-D Fokker-Planck equation for $f(X,\tau)$ reduces to solving two linear algebraic problems:  Eq.~(\ref{geneq:trace}) and Eq.~(\ref{geneq:a_sol}).

\subsection{Application to the $\text{SI}_{\text{1}}\text{I}_{\text{2}}\text{S}$ and  $\text{SI}_{\text{1}}\text{I}_{\text{2}}\text{R}$  models}
\label{sec:application}
We first present the adjoints of operators $L^{(0)}$.  In the $\text{SI}_{\text{1}}\text{I}_{\text{2}}\text{S}$ model, the adjoint of $\hat{L}^{(0)}$ is
\begin{eqnarray}
\label{eq:L0dag2d}
\hat{L}^{(0)\dag} =  h(X)\left(\frac{1}{2}\frac{\partial^2 }{\partial Y^2} - \frac{\mathcal{R}}{2} Y \frac{\partial}{\partial Y}\right).
\end{eqnarray}
In the $\text{SI}_{\text{1}}\text{I}_{\text{2}}\text{R}$ model, the adjoint of $\hat{L}^{(0)}$ is
\begin{equation}
\label{eq:L0dag}
\hat{L}^{(0)\dag} = -\mathcal{R}d_0(X) Z \partial_Y - (\mu \mathcal{R} Z + Y)\partial_Z + c_0(X) \partial^2_{Y} + \mu \partial^2_{Z} + d_0(X) \partial^2_{YZ}.
\end{equation}
These operators are obtained by performing integrations by parts where, because of the Gaussian term, the corresponding functions and their derivatives vanish at the limits of the $Y$-integration.  As the Gaussian is sharp at large $N$, the limits of integration can be extended to $\pm \infty$.

We now calculate the function $F(X,Y)$ described in the previous subsection of this Appendix.  For the $\text{SI}_{\text{1}}\text{I}_{\text{2}}\text{S}$  model we have $\Xi(X)=(\mathcal{R}/2) [(1-a)r+(1+a)X]$,  and Eq.~(\ref{geneq:F}) for the function $F(X,Y)$ becomes
\begin{eqnarray}
\label{geneq:F2}
&&\frac{h(X)}{2}\left(\frac{\partial^2 F}{\partial Y^2}-
\mathcal{R} Y \frac{\partial F}{\partial Y}\right) =  \frac{\mathcal{R} g(X)}{2} Y,
\end{eqnarray}
where $h(X)$ and $g(X)$ are defined in Eqs.~(\ref{eq:nextorder10}) and (\ref{eq:Fsol}), respectively.
The forced solution is readily found:
\begin{equation}
\label{geneq:Fsol}
F(X,Y)=-\frac{g(X)}{h(X)}Y.
\end{equation}
For the $\text{SI}_{\text{1}}\text{I}_{\text{2}}\text{R}$  model Eq.~(\ref{geneq:F}) for the function $F(X,Y,Z)$ becomes
\begin{eqnarray}
&&\hat{L}^{(0) \dag}F(X,Y,Z) =\Xi(X) Z, \\
&&\Xi(X)=\frac{\mathcal{R}a \left[(1-a)r - (1+a)X\right]}{1+a^2},
\end{eqnarray}
where the operator $\hat{L}^{(0)\dag}$ is given in Eq.~(\ref{eq:L0dag}). The forced solution
is
\begin{equation}
\label{geneq:F3d}
F (X,Y,Z)= -\frac{\Xi(X)}{\mathcal{R} d_0(X)}Y, 
\end{equation}
where $d_0(X)$ is given in Eq.~(\ref{eq:c-d-coeffs}).

The operator $\Op L^{(1)\dagger}_Y$ appears in the last line in Eq.~(\ref{geneq:adjoint_structureb}).  Starting with
\begin{equation}
\label{eq:IBP}
 \int_{-\infty}^{\infty} F(X,Y) \left[\Op L^{(1)}\rho^{(0)}(X,Y)\right] dY,
 \end{equation}
we use the explicit form of $\Op L^{(1)}$ from Eq.~(\ref{eq:L12ddefined}) and integrate by parts with respect to $Y$ to arrive at
 \begin{eqnarray}
&-&\frac{(1+a)\mathcal{R}}{2} \int_{-\infty}^{\infty} \frac{\partial F}{\partial Y}Y^2 \rho^{(0)} \,dY +  \int_{-\infty}^{\infty}\frac{\mathcal{R}FY}{2}\frac{\partial }{\partial X}\left[g(X)\rho^{(0)}\right] \,dY  \nonumber \\
 &-& \int_{-\infty}^{\infty} \frac{\partial F}{\partial Y} \frac{\partial }{\partial X}\left[g(X)\rho^{(0)}\right] \,dY +  \int_{-\infty}^{\infty} \frac{1}{2} \frac{\partial^2 F}{\partial Y^2} YH(x) \rho^{(0)} \,dY \nonumber \\
&\equiv&  \int_{-\infty}^{\infty} \left[\Op L^{(1)\dagger}_Y F\right] \rho^{(0)}\,dY\,.
\label{rrr}
 \end{eqnarray}
This should be compared with the last line of Eq.~(\ref{geneq:adjoint_structureb}).  Using the solution for $F$ given by Eq.~(\ref{geneq:Fsol}), we can simplify the r.h.s. of Eq.~(\ref{rrr}) to
\begin{equation}
\label{eq:2dL1result}
\frac{(1+a)g(X) f(X)}{h(X)} + \frac{g^2(X)}{2h(X)} \frac{\partial f}{\partial X},
\end{equation}
which appears in Eq.~(\ref{term2a}).  The calculation for the $\text{SI}_{\text{1}}\text{I}_{\text{2}}\text{R}$ model goes along the same lines: integration by parts can be again applied to Eq.~(\ref{eq:IBP}) to derive $\Op L^{(1)\dagger}_Y$.  Upon substitution of the expression for $F$ from Eq.~(\ref{geneq:F3d}), one arrives at  Eq.~(\ref{eq:adjoint_structureb2}).

\section{Mean time to fixation: analytic results}
\label{appfix}
Evaluating the integrals in Eq.~(\ref{eq:Tgeneral}) and rescaling by $rN$ (everywhere in this Appendix), we obtain for the $\text{SI}_{\text{1}}\text{I}_{\text{2}}\text{S}$ model:
\small
\begin{multline}
\label{eq:T2d}
\!\!\!\!\!T \!=\!\frac{(a-1)^2 \left(\xi_0^2-1\right)-(a+1) (\xi_0-1) [(a-1) \xi_0-a-3] \ln(1-\xi_0)+(a+1) (\xi_0+1) [(a-1) \xi_0 -3a-1] \ln(\xi_0+1)+4 (a+1)^2 \ln 2}{8 a (a+1)}.
\end{multline}
\normalsize
An example of the $\xi_0$ dependence of $T$  is shown in the upper panel of Fig. \ref{T}.  For $a=1$,
\begin{equation}
\label{simpleMTE}
T(\xi_0)=\frac{(\xi_0-1) \ln{(1-\xi_0)}-(\xi_0+1)\ln{(\xi_0+1)} +2\ln{2}}{2}
\end{equation}
for both models, which is symmetric about $\xi_0 = 0$.  As expected, the MTF develops an asymmetry about $\xi_0=0$ as $a$ deviates from $1$.  It also grows as $a$ decreases below $1$.   As $a\to 0$, the dependence on $\xi_0$ and $a$ becomes separable:
\begin{equation}
\label{separable}
T(\xi_0,a\to 0)=\frac{1}{8 a}\left[\xi_0^2+(\xi_0^2+2\xi_0-3) \ln(1-\xi_0)-(1+\xi_0)^2\ln(1+\xi_0)+4\ln2-1\right]\,.
\end{equation}
It has a maximum at $\xi_0 = (e-1)/(e+1)= 0.4621\ldots$.

For the $\text{SI}_{\text{1}}\text{I}_{\text{2}}\text{R}$ model, the expression for $T(\xi_0)$ is very cumbersome.  A relatively simple asymptotic is available for $a\rightarrow 0$:
\begin{equation}
\label{eq:Tapprox3d}
T(\xi_0,a\to 0) \simeq \frac{(\xi_0+1) e^{-\frac{1-\xi_0}{2 a}} \left[\text{Ei}\left(\frac{1-\xi_0}{2 a}\right)+\ln a-\gamma +1\right]-1-\xi_0-2
   \ln (1-\xi_0)+\ln 4}{2 a},
\end{equation}
where $\text{Ei}(x)$ is the exponential integral function and $\gamma=0.5772\ldots $ is the Euler's constant.  The maximum of $T(\xi_0,a\to 0)$ is at the point $\xi_0=\xi_{0\text{max}}$ that satisfies the equation
\begin{equation}
\text{Ei}\left(\frac{1-\xi_0}{2 a}\right)+\ln a=\gamma -1.
\end{equation}
As $a\to 0$, we can drop $\gamma-1$ compared with $\ln a$, and use the large-argument asymptotic $\text{Ei} (w\gg 1) = w^{-1}\,e^w +\ldots$.
This leads to
\begin{equation}
\label{maxnew}
1-\xi_{0\text{max}} = 2 a \ln \ln \frac{1}{a}+ 2 a \ln \ln \ln \frac{1}{a}+\ldots .
\end{equation}
The applicability criterion for Eq.~(\ref{maxnew}) is very stringent: $\ln \ln (1/a)\gg 1$. In the region of $1-\xi_0 \gg 2a$, which includes the maximum point, Eq.~(\ref{eq:Tapprox3d}) simplifies to
\begin{equation}\label{Tcorr1}
T(\xi_0,a\to 0)\simeq \frac{(\xi_0+1) e^{-\frac{1-\xi_0}{2 a}} \ln a-1-\xi_0-2
   \ln (1-\xi_0)+\ln 4}{2 a}.
\end{equation}
In the region of $1-\xi_0 \ll 2a$ we obtain
\begin{equation}\label{Tcorr2}
T(\xi_0,a\to 0)\simeq \frac{(1-\xi_0) \ln
   \left(\frac{2}{1-\xi_0}\right)}
   {2 a^2}
\end{equation}
Interestingly, sufficiently far from the maximum point, $T(\xi_0,a\to 0)$ for $\text{SI}_{\text{1}}\text{I}_{\text{2}}\text{R}$ model shows separability similar to that for the  $\text{SI}_{\text{1}}\text{I}_{\text{2}}\text{S}$ model. To the right of the maximum
point this is evident from Eq.~(\ref{Tcorr2}). To the left of the maximum point the separability emerges when one neglects
the first term in the numerator of Eq.~(\ref{Tcorr1}).  The separability breaks down in the region of maximum. Finally, the maximum value of $T$ can be roughly estimated as
\begin{equation}
\label{eq:Tmaxscaling}
T_{max}(a\to 0) \sim \frac{1}{a} \ln\frac{1}{a}.
\end{equation}

\twocolumngrid


\begin{thebibliography}{99}
\bibitem{comp1} J. P. Grover, \textit{Resource Competition} (Population and Community Biology Series), (Chapman and Hall, Springer, 1997).
\bibitem{comp2} K. J. Rothman and S. Greenland, \textit{Modern Epidemiology}, (Lippincott, Williams \& Wilkins, Philadelphia, 1998).
\bibitem{comp3} E. Tielkes, \textit{Competition for Resources in a Changing World: New Drive for Rural Development} (Cuvillier, G\"{o}ttingen, 2008).
\bibitem{lasers} A. E. Siegman, \textit{Lasers} (University Science Books, Herndon, 1986).
\bibitem{Ostwald} W. Ostwald, Z. Phys. Chem 37, 585 (1901); I. M. Lifshitz and V. V. Slyozov, Sov. Phys. Solid State
1, 1285 (1960); J. Phys. Chem. Solids 19, 35 (1961); C. Wagner, Z. Elektrochem. 65, 581 (1961); L. Ratke and P. Voorhees, \textit{Growth and Coarsening: Ostwald
Ripening in Material Processing} (Springer, New
York, 2002).

\bibitem{Keeling} M. J. Keeling and P. Rohani, \textit{Modeling Infectious Diseases in Humans and Animals} (Princeton University Press, Princeton and Oxford, 2008);

\bibitem{Balmer} O. Balmer and M. Tanner, Lancet Infect Dis. \textbf{11}, 868 (2011).

\bibitem{SISandSI} M.S. Bartlett, \textit{Stochastic Population Models in Ecology
and Epidemiology} (Wiley, New York, 1961); N. T. J. Bailey,
\textit{The Mathematical Theory of Infectious Diseases and its Applications},
(Grifin, London, 1975); R.M. Anderson, and R.M. May,
\textit{Infectious Diseases of Humans: Dynamics and Control}
(Oxford University Press, Oxford, 1991).

\bibitem{Brauer} F. Brauer, P. van den Driessche, J. W, \textit{Mathematical Epidemiology} (Springer-Verlag, Berlin, Heidelberg, 2008).

\bibitem{Parsons1} T. L. Parsons and C. Quince, Theor. Popul. Biol. \textbf{72}, 468 (2007).

\bibitem{Karrer} B. Karrer and M. E. J. Newman, Phys. Rev. E \textbf{84}, 036106 (2011).

\bibitem{Grassberger} L. Chen, F. Ghanbarnejad, W. Cai, and P. Grassberger, EPL \textbf{104}, 50001 (2013).

\bibitem{Parsons2} T. L. Parsons, C. Quince, and J. Plotkin, Theor. Popul. Biol. \textbf{74}, 302 (2008).

\bibitem{Doering} Y. T. Lin, H. Kim, and C. R. Doering, J. Stat. Phys. 148, 646 (2012).

\bibitem{removed} There is also an equation for the recovered or removed population, but it is decoupled from the
rest of equations (\ref{eq:MF-original1}) and therefore unnecessary for our purposes.

\bibitem{Gillespie} D. T.  Gillespie, J. Phys. Chem. \textbf{81}, 2340-2361 (1977).

\bibitem{Gardiner} C. W. Gardiner, \textit{Handbook of Stochastic Methods} (Springer, Berlin, 2004).

\bibitem{SISextinction} 
I. N{\aa}sell, Adv. Appl. Probab. \textbf{28}, 895 (1996); H. Andersson and B. Djehiche, \textit{ibid} \textbf{35}, 662
(1998); I. N{\aa}sell, J. Theor. Biol. \textbf{211}, 11 (2001); O. Ovaskainen, J. Appl. Probab. \textbf{38}, 898 (2001); C. R. Doering, K. V. Sargsyan, and L. M. Sander, Multiscale Model. Simul. \textbf{3}, 283 (2005); M.I. Dykman, I.B. Schwartz and A. S. Landsman, Phys. Rev. Lett. \textbf{101}, 078101 (2008); I. B. Schwartz, L. Billings, M. Dykman and
A. Landsman, J. Stat. Mech. (2009), P01005; M. Khasin and M. I. Dykman, Phys. Rev. Lett. \textbf{103}, 068101
(2009); M. Assaf and B. Meerson, Phys. Rev. E \textbf{81}, 021116 (2010); M. Khasin, B. Meerson, and P. V. Sasorov, Phys. Rev. E \textbf{81}, 031126 (2010).

\bibitem{KM} A. Kamenev and B. Meerson, Phys. Rev. E \textbf{77}, 061107 (2008).

\bibitem{SIextinction} O. A. van Herwaarden and J. Grasman, J. Math. Biol. \textbf{33}, 581 (1995); I. N{\aa}sell, J. R. Stat. Soc. Ser. B. (Stat. Methodol.) \textbf{61}, 309 (1999).

\bibitem{BlytheMcKane} R.A. Blythe and A. McKane, J. Stat. Mech. (2007) P07018.

\bibitem{Born} M. Born and J. R. Oppenheimer, Ann. der Physik \textbf{389}, 457 (1927).

\bibitem{phasediffusion} S.M. Rytov, Zh. Eksp. Teor. Phys. \textbf{29}, 304 (1955) [Sov. Phys. JETP \textbf{2}, 217 (1955)]; M. Lax, in \emph{Statistical Physics, Phase Transisions and Superconductivity}, edited by M. Chretien, E.P. Gross, and S. Deser (Gordon and Breach, New York, 1968).

\bibitem{phasedrift} A. Demir, A. Mehrotra, and J. Roychowdhury, IEEE Trans. Circ. Syst. Fund. Theor. App. \textbf{47}, 655 (2000); H. Nakao, J. Teramae, D. S. Goldobin, and Y. Kuramoto, Chaos \textbf{20}, 033126 (2010).

\bibitem{PK} M. Parker and A. Kamenev, Phys. Rev. E \textbf{80}, 021129 (2009); J. Stat. Phys. \textbf{141}, 201 (2010).

\bibitem{MSK} B. Meerson, P.V. Sasorov, and Y. Kaplan,   Phys. Rev. E \textbf{84} 011147 (2011).

\bibitem{KhM} E. Khain and B. Meerson, J. Phys. A: Math. Theor. \textbf{46}, 125002 (2013); \textit{ibid} \textbf{46} 125002 (2013).

\bibitem{KS} D.A. Kessler and L.M. Sander, Phys. Rev. E \textbf{80}, 041907 (2009).


\end{thebibliography}
\end{document}